\documentclass[fleqn,usenatbib]{mnras}

\usepackage{newtxtext,newtxmath}

\usepackage[T1]{fontenc}

\DeclareRobustCommand{\VAN}[3]{#2}
\let\VANthebibliography\thebibliography
\def\thebibliography{\DeclareRobustCommand{\VAN}[3]{##3}\VANthebibliography}

\usepackage{color,soul}  
\usepackage{natbib}
\usepackage{subcaption}
\usepackage{xcolor}
\usepackage{graphicx}	
\usepackage{amsmath}	
\usepackage{hyperref}

\title[Material Transport in Protoplanetary Discs]{Material Transport in Protoplanetary Discs with Massive Embedded Planets}

\author[H. J. Petrovic et al.]{
Hannah J. Petrovic,$^{1}$\thanks{E-mail: hjp46@cam.ac.uk}
Richard A. Booth,$^{2,3}$
Cathie J. Clarke$^{1}$
\\
$^{1}$Institute of Astronomy, University of Cambridge, Madingley Road, Cambridge CB3 0HA, UK\\
$^{2}$School of Physics and Astronomy, University of Leeds, Leeds, LS2 9JT\\
$^{3}$Astrophysics Group, Imperial College London, Blackett Laboratory, Prince Consort Road, London, SW7 2AZ, UK\\
}

\date{Accepted 2024 September 24. Received 2024 September 24; in original form 2024 August 5}

\pubyear{2024}

\begin{document}
\label{firstpage}
\pagerange{\pageref{firstpage}--\pageref{lastpage}}
\maketitle

\begin{abstract}
Vertical gas and dust flows in protoplanetary discs waft material above the midplane region in the presence of a protoplanet. This motion may alter the delivery of dust to the planet and its circumplanetary disc, as well as through a planetary-induced gap region and hence the inner disc chemistry. Here, we investigate the impact of a massive embedded planet on this material transport through the gap region. We use 3D global hydrodynamic simulations run using FARGO3D with gas and dust species to investigate the dust filtration and the origin of material that can make it through the gap. We find small dust particles can pass through the gap as expected from results in 2D, and that this can be considered in two parts - filtering due to the planetary-induced pressure maximum, and filtering due to accretion onto the planet. When gas accretion onto the planet is included, we find that the larger dust grains that cross the gap (i.e. those with $\mathrm{St} \sim 10^{-4}$) originate from regions near the mid-plane. We also find that dust and gas that enter the planet-carved gap region pass through the Hill sphere of the planet, where the temperature is likely to be strongly enhanced compared with the mid-plane regions from which this material originated. Considering the application of our simulations to a Jupiter-mass planet at $\sim 100\ \mathrm{AU}$, this suggests that CO ice is very likely to desorb from grains in the close proximity of the planet, without requiring any fine-tuning of the planet's location with respect to the CO snowline.

\end{abstract}



\section{Introduction}\label{Sect 1: Introduction}

Recent confirmations of protoplanet candidates such as PDS 70 b and c with mass estimates of several Jupiter masses each \citep{Keppler2018, Haffert2019, Mesa2019} and their associated rings, continue to strengthen the theory that protoplanets can form structure in protoplanetary discs (PPDs). Understanding material transport to the planet and through the disc in the presence of massive embedded planets such as these then becomes crucial to understanding what consequences this has on the disc and planetary formation. Not only does studying this motion allow us to comment on the origin of material that comprises the inner disc regions available for further planetary growth, but it also provides the first steps in understanding the processes this material experiences on its journey to the inner disc region, important for understanding the complex chemistry present in terrestrial planets.

It is well known that the presence of a massive embedded planet can lead to the formation of a gas gap in the disc \citep{Lin1986} which subsequently inhibits the movement of large dust grains through the gap region as the dust becomes stuck at the planet-induced pressure maximum \citep{Paardekooper2004, Paardekooper2006}. This filtering is well studied in 2D where it has been  shown that dust with Stokes numbers less than the prescribed alpha viscosity \citep{Shakura1973} in the disc, $\mathrm{St} < \alpha$, are able to flow through a gap \citep{Rice2006}. However, 2D simulations are perhaps no longer sufficient to model this behaviour. Studies have shown there are significant vertical gas flows in these discs, including the meridional gas flows that transport material into the circumplanetary disc \citep{Tanigawa2012, Morbidelli2014Apr, Szulagyi2014}, and these are  likely to change the flow of dust through the planetary-induced gap region. 

As seen in pre-ALMA theoretical studies, vertical settling of dust grains is also an important factor for dust evolution in these discs, implying that mm-sized dust strongly settles to the mid-plane while smaller $\mu$m-sized dust is present over several scale heights \citep{Dubrulle1995, Takeuchi2002, Fouchet2007}. This has since been verified in ALMA observations of edge-on discs showing this mm-sized dust settled in the mid-plane \citep{Villenave2020}. However, \citet{Bi2021} show that massive planets can stir small dust grains to high altitudes near the gap edges, thought to be due to the planet-induced meridional flows. \citet{Szulagyi2022} and \citet{Karlin2023} support this, showing that accretion into the circumplanetary disc region occurs through the poles, with outflow of material away from the planet in the mid-plane. In each case, these papers present facets of the gas and dust motion that support the need for 3D simulations to most accurately model their behaviour in these discs. 

Recent studies have also proposed a correlation between super-Earths in systems with a cold Jupiter \citep{Zhu2018, Bryan2019, Rosenthal2022, Bryan2024} suggesting that a formation pathway for these planets is needed, while the formation of a Jupiter-mass planet is typically thought to disrupt the availability of material inside its orbit due to trapping. While this is still under debate \citep{Barbato2018, Schlecker2021, Bonomo2023}, it is important to understand the impact of a giant planet on the flow of material through the disc as a result. The presence of an inner dust disc inside the radial locations of PDS 70 b and c further supports the idea that material can pass through a planetary-induced gap region from the outer disc into the inner disc region \citep{Pinilla2024}. Although \citet{Best2024} suggest transport inwards in discs with massive embedded planets could be explained by secular resonance sweeping that can move planetesimal rings inwards by increasing their eccentricities, subsequently enhancing the planetesimal surface density in the inner disc, it is important to consider the motion of smaller dust through the disc that could address this problem.

In this paper we conduct 3D global hydrodynamic simulations of gas and dust in protoplanetary discs with an embedded planet with planet to star mass ratio of $10^{-3}$ and a disc aspect ratio $(h/r)$ of 0.1 at the planet location. This latter choice is motivated by the increasingly stringent resolution requirements for thinner discs and the fact that our long timescale integrations (thousands of orbits at the planet location) are computationally expensive. Such a choice means that the calculations are most directly comparable to the case of planets located at large radii in the disc (30-100 AU). For a solar mass star, the planet mass is close to one Jupiter mass and corresponds, for this disc aspect ratio, to a planet mass that is twice the pebble isolation mass. The pebble isolation mass describes the minimum mass of an embedded planet required to create a pressure maximum in the disc that can inhibit the movement of typically mm-sized grains through a gap and their accretion onto the planet \citep{Lambrechts2014a}. We explore the implications of dust filtering due to the planet-induced pressure maximum, investigating which dust species are able to permeate through the gap region, considering a range of simulation parameters. Through tracking of particle paths we study the origin of material that can reach the inner disc, potentially being available for further planet building in the inner disc. 

The paper is organised as follows: in Section \ref{Sect 2: Numerical Setup} we describe the numerical setup in our simulations, including the accretion prescription applied, the simulation parameters, and a description of the Lagrangian particle tracking method used to further investigate the motion through the disc. In Section \ref{Sect 3: Results} we present the results. We consider the implications of these results and include a discussion in Section \ref{Sect 4: Discussion} with a focus on the significance of the paths taken for material that is able to make it through the gap within our evolved discs, before concluding in Section \ref{Sect 5: Conclusion}.

\section{Methods}\label{Sect 2: Numerical Setup}

Here we describe the simulation setup and parameters used, along with the accretion algorithm added to compute accretion onto the planet. We detail the simulation parameters used and a description of additional test particle tracking from the outputs of these simulations.

\subsection{Numerical Model}
\subsubsection{Simulation Setup}

We use the hydrodynamics code FARGO3D \citep{Benitez-Llambay2016} including treatment of dust dynamics \citep{Rosotti2016} with the orbital advection algorithm \citep{Masset2000}. The accretion prescription described later in Section~\ref{Sect 2: Accretion Prescription} has been added in this work to include the accretion of gas and dust onto the planet. Dimensionless units are used throughout such that $G = M_\star = r_0 = 1$, where $G$ is the gravitational constant, $M_\star$ is the mass of the central star, and $r_0$ is the fixed radial orbit of the planet, $r_0 = r_p = 1$.

FARGO3D solves the hydrodynamic equations, having first subtracted the Keplerian velocity field, on a fixed mesh. The equations solved are the continuity and momentum equations,
\begin{equation}
    \frac{\partial \rho}{\partial t} + \nabla\cdot (\rho \mathbf{v}) = 0,
\end{equation}
\begin{equation}
    \frac{\partial \mathbf{v}}{\partial t} + \mathbf{v}\cdot \nabla\mathbf{v} + \frac{\nabla P}{\rho} = \mathbf{g},
\end{equation}

where $\mathbf{v}$ is the velocity in a cell, $\rho$ is the density, $P$ the pressure and $\mathbf{g}$ a gravity term. The Navier-Stokes equation,
\begin{equation}
    \begin{aligned}
        \rho\bigg(\frac{\partial \mathbf{v}}{\partial t} 
        + \mathbf{v}\cdot\nabla\mathbf{v} \bigg) 
        &= -\nabla P
        + \nabla\cdot\mathbf{T} 
        + \mathbf{F}_\mathrm{ext} \\ 
        &- [2\boldsymbol{\Omega} \times\mathbf{v} 
        + \Omega \times (\boldsymbol{\Omega} \times \mathbf{r}) 
        + \dot{\boldsymbol{\Omega}} \times \mathbf{r}]\rho,\\
    \end{aligned} 
\end{equation}
is also solved where $\mathbf{F}_\mathrm{ext}$ is an external force due to gravity and $\mathbf{T}$ is the viscous stress tensor. A pressure-density relation for a locally isothermal equation of state with the form $P=\rho c_s^2$ is used, where $c_s$ is the local sound speed. No magnetic fields are included. 

Running these simulations in 3D, reflection symmetry about the mid-plane is assumed such that only one side of the disc is modelled using spherical coordinates $(r,\theta,\phi)$. In post-processing, cylindrical coordinates in $(\phi,R,z)$ are also used to show the results. We use a domain of $r \in [0.3,3.0]$ in radius, with cells spaced logarithmically, a polar extent of $\theta\in [\pi/2 - 3H_g,\pi/2]$, where $\theta$ is defined from the axis of symmetry and $H_g$ is the scale height of gas in the disc at the planet location, and $\phi\in [-\pi,\pi]$ in azimuth. Both the azimuthal and polar cells are spaced uniformly, and we use code units throughout. A resolution of $(N_r, N_\theta, N_\phi) = (512,60,1024)$ is chosen resolving the scale height in the disc polar direction by 20 cells at the planet radial location.

Closed boundaries are used for the radial boundaries and for the surface of the disc to prevent additional inflow of material. In particular, re-supply of dust through the outer radial boundary is prevented in this way in order to allow us to investigate the differences between motion of various dust species into the inner disc as we initialise each dust species with the same initial mass in the disc. This lack of re-supply from large radii means that at any subsequent time in the simulations, this fixed initial dust budget is divided between fractions that are a) still in the disc, and b) accreted by the planet. The mid-plane boundary is reflecting to mimic the full disc behaviour in our half disc. Stockholm damping is prescribed for the radial velocity of the gas in the inner and outer boundary regions to prevent the production of artifacts from reflection at the boundaries \citep{deVal-Borro2006, McNally2019}. This damping is not applied for the dust species. 

The disc itself is modelled using a flaring index of $f = 0.25$, where the disc aspect ratio is described by $h = h_0 r^{f}$, where $h_0$ is the aspect ratio initialised at the planet location, $r_p = 1$. The aspect ratio is also described by $h = H/r$ with $H$, the disc scale height, defined by $H = c_s/\Omega$, where the sound speed $c_s$ is a function of cylindrical radius $R$. Here we use a disc with $H/R = 0.1$. The flaring increases the disc thickness with distance from the star \citep{Kenyon1987}, such that the temperature profile follows,
\begin{equation}
    T(R) \propto \bigg(\frac{R}{R_0}\bigg)^{-1/2}.
\end{equation}
 
From \citet{Shakura1973} the kinematic viscosity is described as,
\begin{equation}
    \nu = \alpha c_s H,
\end{equation}
using the alpha viscosity, $\alpha$. For our fiducial setup this is set to $\alpha = 10^{-3}$, however, we include simulations with larger alpha values for comparison. Dust diffusion is also included with a default Schmidt number (ratio of kinematic viscosity to diffusivity) of $\mathrm{Sc} = 1.0$ unless stated otherwise.
Diffusion is applied by modifying the speed at which the dust density and momentum are advected at, using diffusive velocities computed with $\mathbf{v_D} = \mathbf{F_D}/\Sigma_d$, where $\mathbf{F_D}$ describes the diffusive mass flux as $\mathbf{F_D} = -(\nu/\mathrm{Sc}) \Sigma_g\nabla(\Sigma_d/\Sigma_g)$. Here $\Sigma$ describes the surface density of the dust ($d$) or the gas ($g$). A weaker dust diffusion using a value of $\mathrm{Sc} = 10.0$ is also investigated. 

The initial density profile in the gas follows 
\begin{equation}
    \rho_{g0} = \rho_{g0,\rm{ref}}\Bigg( \frac{R}{R_{\rm{ref}}} \Bigg)^{p} \times 
    \exp \Bigg[\frac{GM_\star}{{c_s}^2} \Bigg(\frac{1}{r} - \frac{1}{R}\Bigg) \Bigg],
\end{equation}
such that the density in the gas has an approximately Gaussian profile in the vertical extent of the disc, and a power law dependence on cylindrical radius, $R$. Here $p = -1$ is adopted and $\rho_{g0,\rm{ref}}$ is related to the surface density, at the reference value $r_0 = R_\mathrm{ref} = 1$ according to,
\begin{equation}\label{Eqn: Initial Density Profile}
    \rho_{g0,\rm{ref}} = \frac{\Sigma_{g0}}{\sqrt{2\pi} H}.
\end{equation} 
The value of $\Sigma_{g0}$ is $6.3\times10^{-4}$ in code units, resulting in a gas disc mass of $M_g = 0.01M_\star$, in the case that $r=1$ corresponds to 100 AU. 

We investigated the transport and accretion of grain sizes with mid-plane Stokes numbers at the planet location between $10^{-5} - 10^{-1}$. The initial dust density is assumed to follow the above density distribution, adopting an initially fixed dust-to-gas ratio of $0.01$. The choice of this dust-to-gas ratio is arbitrary since we neglect feedback from the dust on the gas in these simulations.

As described in Section~\ref{Sect 1: Introduction}, we focus on a planet mass that exceeds the pebble isolation mass. We use the equivalent of a Jupiter mass planet in our code-unit simulations such that $M_p = 0.001 M_\star$, which is kept fixed. The planet is kept on a fixed circular orbit with a gravitational smoothing parameter of $0.1r_H$, the value of which is the same as for the sink region used in our accretion prescription described in Section~\ref{Sect 2: Accretion Prescription}, so as to avoid smoothing beyond the accreting region. Here $r_H$ is the Hill sphere of the planet, described by,
\begin{equation}\label{Eqn:Hill Radius}
    r_H \equiv r_p \bigg( \frac{M_p}{3M_\star}\bigg)^{\frac{1}{3}}.
\end{equation}

The simulations are allowed to run for a minimum of 1000 planetary orbits for the large dust species that evolve quickly, with longer run times up to 2000 orbits used for the smallest dust grains that take longer to reach a quasi-steady state. This quasi-steady state is determined such that the azimuthally averaged surface profiles show only small variations over several hundred orbits. Since \citet{Bi2021} show that the azimuthally averaged gas profile is still evolving at 1000 orbits in 3D simulations, but the profiles at 2000 and 3000 orbits show good agreement, we run the gas in each case for 2000 orbits and as such suggest this should also be sufficiently long for the well-coupled small dust grains above. Since the total duration of the simulation is a small fraction of the viscous timescale at the location of the planet, one therefore expects the gas to evolve only as a result of the localised changes induced by the planet. The full set of simulations run and their key disc and dust parameters are shown in Table~\ref{Table: sim_parameters}.

\begin{table*}
	\centering
	\caption{Simulation parameters. Note that the name of each simulation used illustrates the key simulation parameters starting with the equivalent to a Jupiter planet mass for a solar mass star, and noting the Schmidt number ($\mathrm{Sc}$) used for dust diffusion, the disc alpha viscosity ($\alpha$), and whether gas accretion is included should these values vary from the fiducial model. The table includes these values of $\alpha$, $\mathrm{Sc}$, the total number of dust species run, and the Stokes numbers of these dust initialised at the planet location at simulation setup. The last column specifies whether gas accretion is included alongside the dust accretion that is included for all simulations.}
    \label{Table: sim_parameters}
	\begin{tabular}{lcccccc} 
		\hline
		Simulation & $\alpha$ & Sc & \# Dust Species & Stokes Numbers & Gas Accretion \\
		\hline
            Fiducial & $10^{-3}$ & 1.0 & 5 & $10^{-1}, 10^{-2}, 10^{-3}, 10^{-4}, 10^{-5}$ & Yes\\
            Jup\_Sc10 & $10^{-3}$ & 10.0 & 4 & $ 10^{-2}, 10^{-3}, 10^{-4}$ & Yes\\
            Jup\_alpha3\_3 & $3\times10^{-3}$ & 1.0 & 5 & $10^{-1}, 10^{-2}, 10^{-3}, 10^{-4}, 10^{-5}$ & Yes\\
            \hline
            Jup\_no\_gasacc & $10^{-3}$ & 1.0 & 8 & $10^{-1}, 10^{-2}, 7\times10^{-3}, 5\times10^{-3}, 3\times10^{-3}, 10^{-3}, 10^{-4}, 10^{-5}$  & No\\
            Jup\_Sc10\_no\_gasacc & $10^{-3}$ & 10.0 & 4 & $ 10^{-2}, 5\times 10^{-3},  10^{-3}, 10^{-4}$ & No\\
            Jup\_alpha2\_no\_gasacc & $10^{-2}$ & 1.0 & 5 & $10^{-1}, 10^{-2}, 10^{-3}, 10^{-4}, 10^{-5}$ & No\\
            Jup\_alpha3\_3\_no\_gasacc & $3\times10^{-3}$ & 1.0 & 5 & $10^{-1}, 10^{-2}, 10^{-3}, 10^{-4}, 10^{-5}$ & No\\
		\hline
	\end{tabular}
\end{table*}

\subsubsection{Dust Dynamics}

We treat the dust as a pressureless fluid such that it would describe Keplerian orbits in the absence of gas in the disc. The presence of gas therefore influences the dynamics of the dust, with the impact of this effect determined by the Stokes number in terms of the stopping time of the particles, $t_s$, as
\begin{equation}
    \mathrm{St} = t_s\Omega_K,
\end{equation}
where the stopping time characterises the drag force experienced by the dust grains as described below. 

We assign fixed particle grain sizes to each dust fluid across the domain. These are specified in terms of the initial mid-plane Stokes number at the location of the planet at $r_p = 1$ as shown in Table~\ref{Table: sim_parameters}. The drag is then modelled assuming the particles are in the Epstein regime, such that in the limit of Epstein drag the Stokes numbers are given by
\begin{equation}
    \mathrm{St} = \frac{a \rho_i \Omega_K}{\rho_g c_s}.
\end{equation}
Here the dust grain internal density, $\rho_i$, is set to a constant value of $\rho_i = 1$ gcm$^{-3}$ to represent approximately micron-sized grains for $\mathrm{St} = 10^{-1}$ initialised at the planet radius in the mid-plane. $a$ denotes the radial size of the grains, and $\rho_g$ the surrounding gas density. The stopping time of these particles is computed through,
\begin{equation}
    t_s = \frac{\rho_\mathrm{g0,ref}}{\rho_g}\frac{c_\mathrm{s,ref}}{c_s}\frac{\mathrm{St_\mathrm{0,ref}}}{\Omega_\mathrm{K,ref}},
\end{equation}
using reference values in the mid-plane at $r_p$ denoted with $\mathrm{ref}$, and is used in the equation for the dust velocity, given by
\begin{equation}
    \frac{\partial \mathbf{v}_d}{\partial t} + \mathbf{v}_d \cdot \nabla \mathbf{v}_d = -\frac{1}{t_s}(\mathbf{v}_d - \mathbf{v}_g) + \mathbf{g},
\end{equation}
where $\mathbf{v}_d$ specifies the dust velocity and $\mathbf{v}_g$ is the gas velocity.

The effect of drag is to cause dust to drift towards a pressure maximum \citep{Whipple1972,Weidenschilling1977}, which is typically the disc inner edge in the case of a smooth disc. Hence, in the case of a massive embedded planet that can alter the gas density, depleting the gas density at its radial location to create a gap, the dust can be stopped beyond the radial orbit of the planet at the planet-induced pressure maximum. The minimum mass for a planet to create a pressure maximum is known as the isolation mass, $M_\mathrm{iso}$, which is approximately given by
\begin{equation}\label{Eqn:Pebble Isolation Mass}
    M_\mathrm{iso} \sim 20 M_{\oplus} \bigg( \frac{H/r}{0.05} \bigg)^3,
\end{equation}
determined using hydrodynamical simulations in 2D in \citet{Lambrechts2014a}, and in 3D in \citet{Bitsch2015}. We use $H/R = 0.1$ throughout, resulting in $M_\mathrm{iso} = 0.5M_J$. For our simulations the planet mass is therefore twice the isolation mass, such that a deep gap similar to those observed is formed.

\subsubsection{Accretion Prescription}\label{Sect 2: Accretion Prescription}

Accretion onto the planet was included following the method from \citet{Li2021} and previously in \citet{Tanigawa2002Nov}, \citet{D_Angelo2003} and \citet{Dobbs_Dixon2007}. This method uses a sink at the planet location from which the mass accretion rate onto the planet is determined via,
\begin{equation}\label{Eqn: AccretionRate}
    \dot{m}_p = \frac{f}{\tau} \int_{0}^{\Delta} 4\pi \rho (r) r^2 dr.
\end{equation}
Here, $f/\tau$ describes the fraction of mass that is accreted per unit time within the sink region, as $f$ describes the accretion factor and $\tau$ the timescale for the removal of this mass. We employ $f/\tau = \frac{5}{\pi}\Omega_p$ for the gas, where $\Omega_p$ is the Keplerian angular frequency of the planet. We used $f/\tau = \frac{20}{\pi}\Omega_p$ for the dust, which was found to be robust (in the sense that the resulting accretion rate being insensitive to the value of $f/\tau$ from the investigation in Appendix \ref{Appendix: Dust Accretion Fraction}). The integral in Eqn.~\ref{Eqn: AccretionRate} is the mass of material in the sink region which has radius $\Delta = 0.1 r_H$.

As giant planets typically have circumplanetary discs (CPDs) of size of $\sim0.3-0.5r_H$ \citep{Quillen1998, Ayliffe2009, Martin2011}, the simulations here include the circumplanetary disc in the computational domain, within the planetary Hill sphere, but it is poorly resolved (see below).

The simulations have a resolution high enough to resolve this sink radius such that at the location of the planet the cells have lengths $dr, rd\theta, rd\phi \approx 0.65\Delta, 0.72\Delta$ and $0.88\Delta$ respectively. The accretion rate is computed by summing the contributions to this value from all cells that lie within the sink region. To prevent overestimation of the accretion rate due to cells with large volumes beyond the sink region, the mass of each cell contributing to accretion onto the planet is adjusted to take into account the proportion of the cell inside the sink region, $P_c$, such that
\begin{equation}
    \dot{m}_p = \frac{f}{\tau} \sum\limits_{\substack{\mathrm{cells} \\ \mathrm{in \, sink}}} P_c V_c \rho_c,
\end{equation}
where $V_c$ is the volume of the cell and $\rho_c$ is its density. Convergence in the accretion rates over different resolutions in 3D can be seen in Appendix~\ref{Appendix: Gas Accretion Resolution Check}.

The accretion prescription is applied to both the gas and dust for three of these simulations, with four simulations run excluding gas accretion. Since it is known that the gas accretion impacts on the gap structure formed \citep{Bergez-Casalou2020, Rosenthal2020}, this allows us to bracket the range of possible accretion rates of the dust, where excluding gas accretion essentially creates an environment where the gas accretes at its slowest, and including gas accretion forms an upper bound on the gas accretion, with potential consequences on the dust accretion rate. 

\subsection{Post-Processing of Particle Trajectories}

We post-processed the simulations using a Lagrangian particle tracking method to investigate the evolution of material once the simulations had reached a quasi-steady state. Both forwards and backwards tracking of particles in time was conducted to check the method works as intended.  

This was conducted by tracking a total of $3120$ particle paths with initial positions evenly distributed along the polar and azimuthal domains at a fixed selected radial location. A time step was chosen such that this was less than the crossing time of a particle across a cell to ensure the movement of a particle during each step was fine enough to provide accurate tracking. When computing the updated locations at each time step, the velocities of particles used are computed by interpolating the velocities in each axis from the FARGO3D outputs. These velocity arrays are averaged over an additional 100 planetary orbits from the end of the simulation since, although the simulation has reached a quasi-steady state by this point such that the average velocities are not changing, there may be fluctuations in these velocity arrays over short timescales captured in individual outputs. This averaging is designed to mitigate the effect of these fluctuations on the computed trajectories. The particles were integrated in a frame co-rotating with the planet, and in each case, particles are allowed to evolve until they reach the location of the damping zone, or until a fixed $t = 4000$ in dynamical time (where the orbital period of the planet is $2\pi$ in these units).

\section{Results}\label{Sect 3: Results}

Initially we study the impact of the planet on the dust filtration into the gap region, before moving on to present a more detailed analysis of particle trajectories through the gap considering starting positions at different heights in the disc.

\subsection{Dust Filtration}\label{Sect 3: Dust Filtration}

\begin{figure}
    \includegraphics[width=\columnwidth]{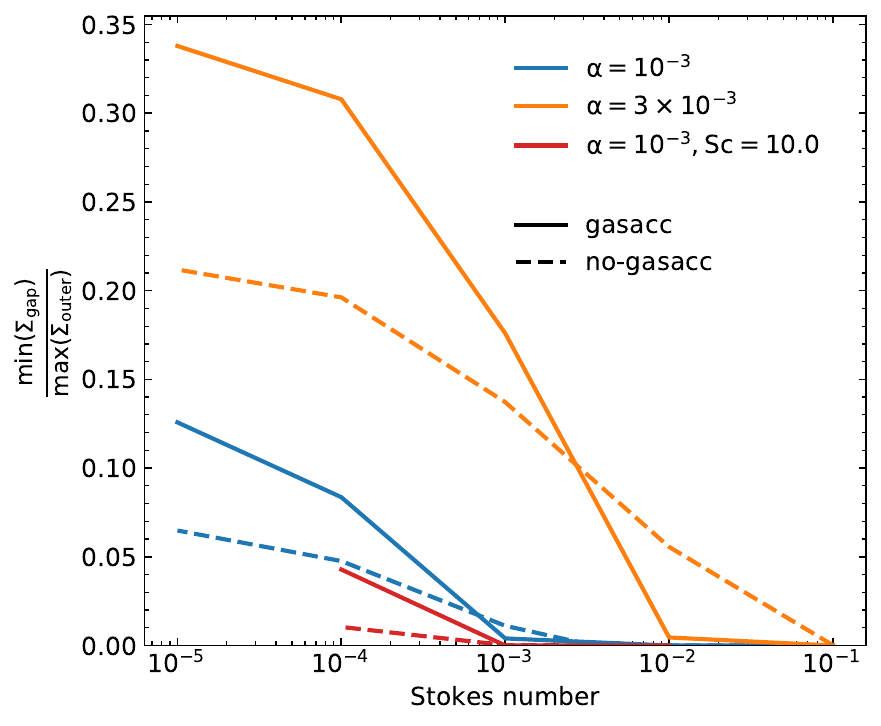}
    \caption{Ratio of dust surface density in the gap (excluding azimuths within $2r_H$ either side of the planet azimuth $\phi_p$) to the maximum in the outer disc as a function of the initial Stokes numbers for the dust in each simulation: simulations with $\alpha = 10^{-3}, \ \mathrm{Sc} = 1.0$ (blue), with $\alpha = 3\times10^{-3}, \ \mathrm{Sc} = 1.0$ (orange), and $\alpha = 10^{-3}, \ \mathrm{Sc} = 10.0$ (red). In each case, a solid line indicates a simulation including gas accretion, while a dashed line indicates a simulation without. The $\alpha = 10^{-2}$ simulation has values close to 1 and are therefore not plotted.}
    \label{fig: Pressure Max Filtration}
\end{figure}

As in the case of previous 2D simulations, we find in our 3D simulations that the dust flux through the gap continues to be dependent on Stokes number, with the well-coupled, i.e., lower Stokes number dust, able to make it through a planetary-induced pressure maximum while larger dust is inhibited in its inwards motion into the gap. This can be seen in Fig.~\ref{fig: Pressure Max Filtration}, where the ratio of the minimum surface density in the gap region between the planet and the outer disc to the maximum surface density in the outer disc is shown. Here it is clear that with increasing grain size (initial Stokes number), the proportion of dust that is prevented moving inwards increases, leading to a deeper gap region. The dust gap depth is also affected by the inclusion (exclusion) of gas accretion, where shallower (deeper) gaps are shown for small grain sizes, while larger dust grains have a deeper (shallower) gap. While this increased filtering of large grains is expected when gas accretion is included due to the deeper gas gaps and therefore filtering on these grains at the pressure maximum, the small grain result is unexpected but arises from a small difference in the gas disc structure, where excluding gas accretion leads to a higher gas surface density at the outer gap edge where the evolution of small grains into the gap region is slowed leading to deeper gaps. We have not included the $\alpha = 10^{-2}$ results here as the larger alpha viscosity prevents the formation of a pressure maximum beyond the planet allowing all particles with $\mathrm{St} \leq 1$ through. 

\begin{figure}
    \includegraphics[width=\columnwidth]{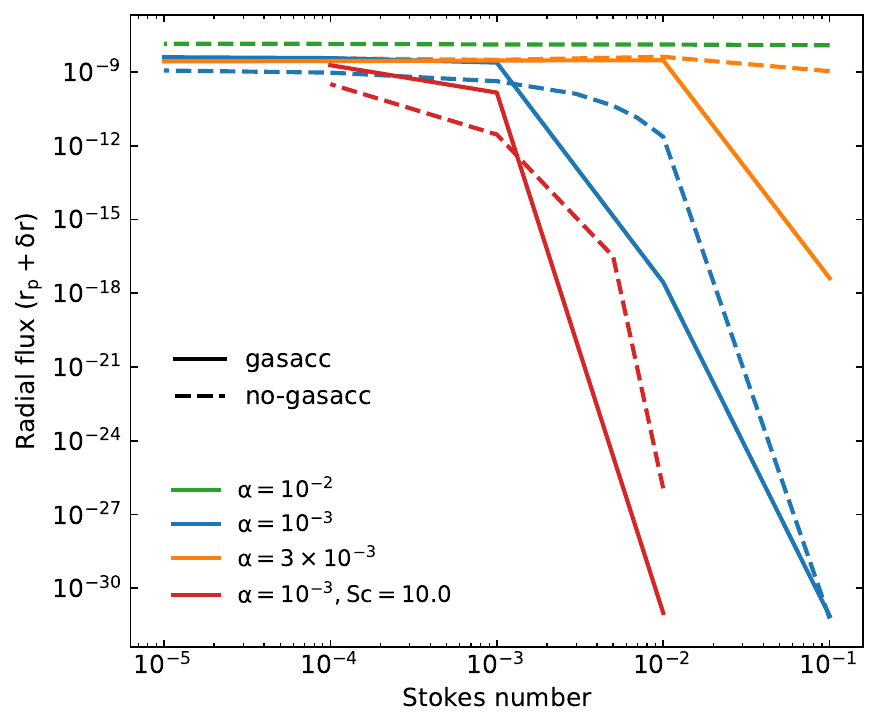}
    \caption{Radial flux of dust in the gap at $r_p + \delta r$ as a function of initial Stokes numbers for the dust in each simulation. Labels are the same as described in Fig.~\ref{fig: Pressure Max Filtration}, with the addition of the $\mathrm{Sc} = 1.0, \alpha = 10^{-2}$ simulation (green).}
    \label{fig: Outer Radial Flux}
\end{figure}

By taking the radial fluxes computed for all cells, and summing over the polar and azimuthal domains, we can show the impact of this pressure-maximum filtering on the radial flux in the gap region. We find that in the fiducial setup, with $\alpha = 10^{-3}$ and $\mathrm{Sc} = 1.0$, the largest grains able to make it into the planetary-induced gap region in non-negligible proportions have initial Stokes numbers of $\mathrm{St} \leq 10^{-3}$ when gas accretion is included, and between $5\times 10^{-3} \leq \mathrm{St} \leq 7\times 10^{-3}$ for the more finely simulated range of dust without gas accretion. This is shown in Fig.~\ref{fig: Outer Radial Flux}, where the radial flux values just beyond the planet location at $r_p + \delta r$ are plotted. Here, $\delta r = 0.02$ such that this radial value is well in the gap region whilst remaining beyond the planet accretion sink. This radial flux includes both the flux that is later accreted by the planet and that which makes it into the inner disc. In each parameter setup with $\mathrm{Sc} = 1.0$, we find that for $\mathrm{St} \leq \alpha$ a significant flux of material is able to make it into the gap region. The $\mathrm{Sc} = 10.0$ simulation shows a similar trend, however, we do not present results for $\mathrm{Sc} = 10.0$ and $\mathrm{St} = 10^{-5}$ due to the prohibitively long timescale required to achieve a steady state.

\begin{figure}
    \includegraphics[width=\columnwidth]{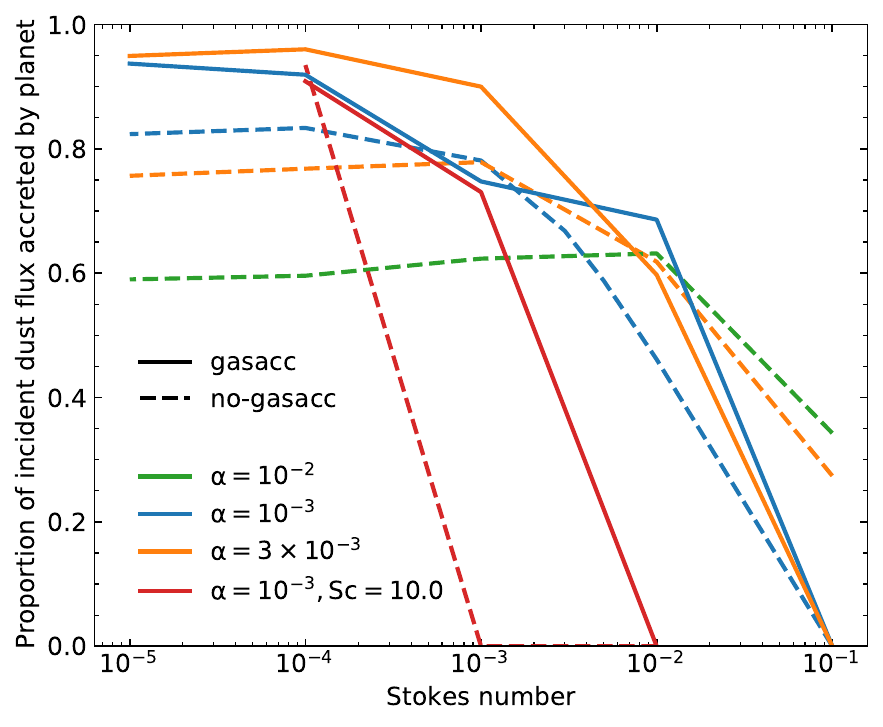}
    \caption{Proportion of dust radial flux reduction due to planetary accretion taken from flux measurements a distance $\delta r$ either side of the planet radially as a function of their initial Stokes numbers. Labels are the same as described in Fig.~\ref{fig: Pressure Max Filtration} and Fig.~\ref{fig: Outer Radial Flux}.}
    \label{fig: Planet Filtering Fraction}
\end{figure}

We can also consider the filtering in the gap region itself due to the accretion onto the planet as seen in Fig.~\ref{fig: Planet Filtering Fraction}. Here the trend for all simulations is that whereas dust with smaller initial Stokes number passes more readily into the gap region (Fig.~\ref{fig: Outer Radial Flux}), it is more likely, once it has entered the gap region, to be accreted by the planet. On the other hand, dust with higher Stokes number is preferentially trapped exterior to the planet; however, grains with higher Stokes number that do succeed in entering the gap are more likely to traverse the gap without being accreted by the planet. This proportion is computed by taking the radial flux, summed over polar and azimuthal domains, at a distance $\delta r$ either side of the planet radially and from this quantifying the proportion unable to pass through the location of the planet. We verify that this difference matches the accretion rate onto the planet computed with our accretion prescription. In the case where material is seen to be stuck beyond the planet from surface density profiles, these values are set to $0.0$. The trends seen in Fig.~\ref{fig: Planet Filtering Fraction} can be explained considering the drift timescales of the dust species - those that are well-coupled to the gas will drift on longer timescales, a consequence of which is spending a longer time in the region from which the planet is able to accrete. Conversely, the larger grains that drift more quickly, have a shorter available time when they are within the accretion region of the planet, and therefore have a reduced accretion fraction and filtration through the planetary location.
This trend is evident regardless of the inclusion of gas accretion. However, it should be noted that when including gas accretion a greater proportion of the small dust that is able to reach the planet is accreted. This, combined with the greater amount of this small dust that can pass through into the gap region, leads to a greater proportion of material accreted by the planet for these species.

\subsection{Meridional Motion}\label{Sect 3: Meridional Motion}

\begin{figure*}
\begin{subfigure}{0.8\textwidth}
  \centering
  \includegraphics[width=\columnwidth]{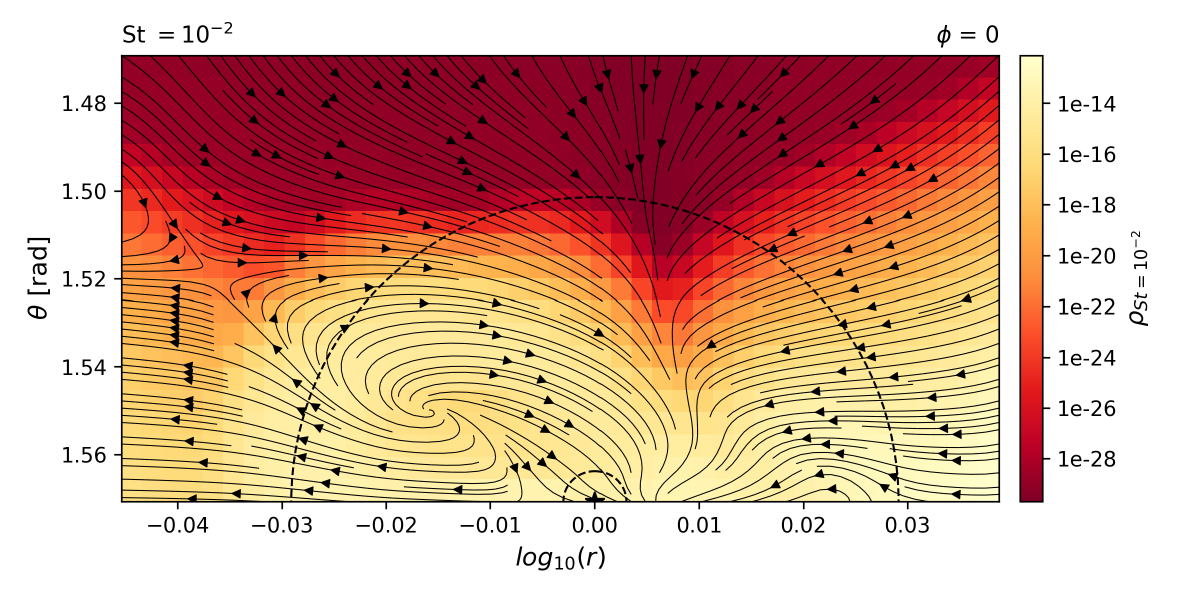}
  \caption{$\mathrm{St} = 10^{-2}$}
  \label{fig: St=10^-2 Meridionals}
\end{subfigure}
\begin{subfigure}{0.8\textwidth}
  \centering
  \includegraphics[width=\columnwidth]{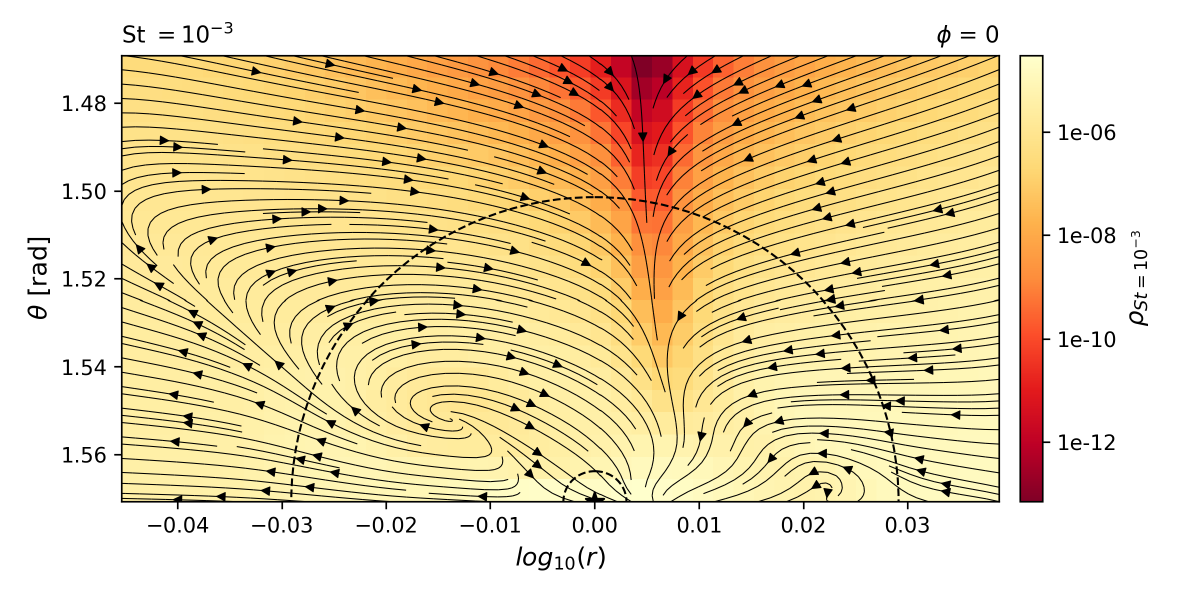}
  \caption{$\mathrm{St} = 10^{-3}$}
  \label{fig: St=10^-3 Meridionals}
\end{subfigure} 
\caption{Streamline plots for dust particles with initial Stokes numbers of $\mathrm{St} = 10^{-2}$ (top) and $\mathrm{St} = 10^{-3}$ (bottom) in the fiducial simulation with gas accretion at 2000 orbits. Arrows show the motion in the $(\mathrm{log}_{10}(r),\theta)$ plane at the azimuthal location of the planet, $\phi = 0.0$, where the planet is denoted by a star symbol, with two surrounding dashed circles for the sink and Hill sphere of the planet respectively. The background colormap shows the density in the slice at the same location for each species. In each case, meridional flows are shown with motion towards the planet vertically, while outwards flows radially from the planet move material away from the planet.}
\label{fig: Meridional Motion}
\end{figure*}

\begin{figure*}
    \centering
    \begin{subfigure}{0.49\textwidth}
        \centering
        \includegraphics[width=\textwidth]{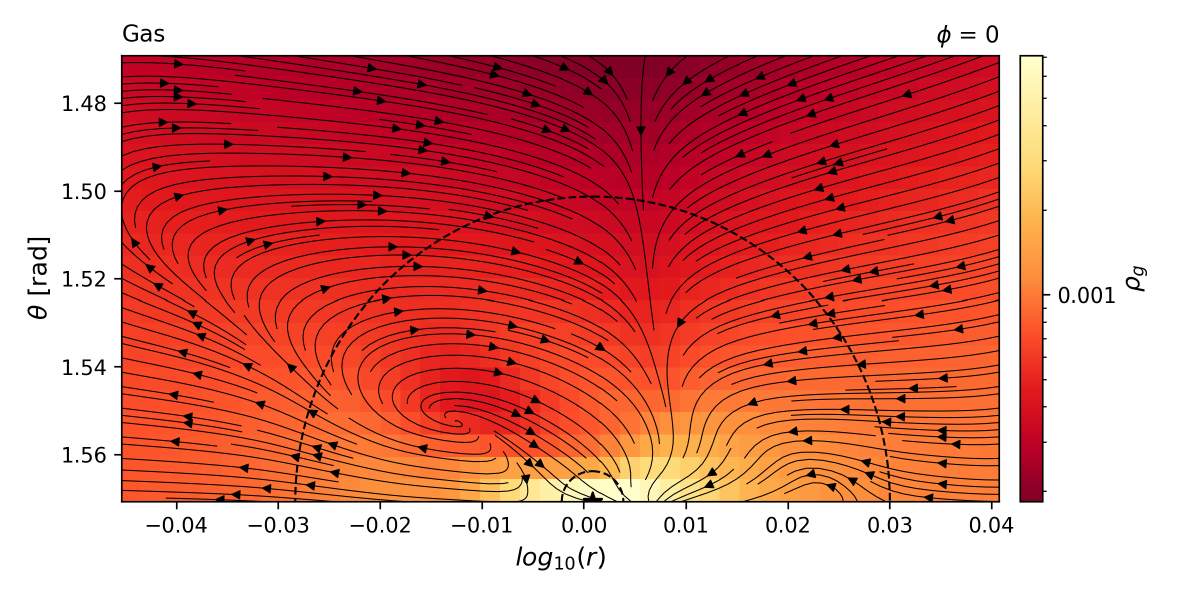}
        \caption{Gas accretion, $t=1000$ orbits}
        \label{fig: Gas acc, 1000 orbits}
    \end{subfigure}
    \begin{subfigure}{0.49\textwidth}
        \centering
        \includegraphics[width=\textwidth]{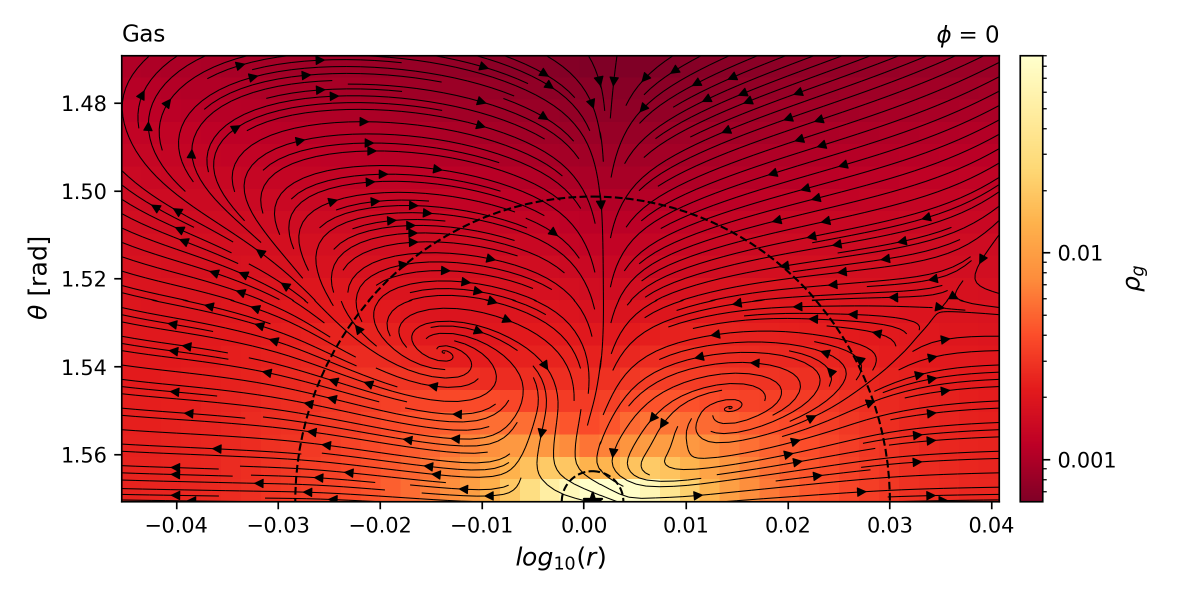}
        \caption{No gas accretion, $t=1000$ orbits}
        \label{fig: No gas acc, 1000 orbits}
    \end{subfigure}
    \vskip\baselineskip
    \begin{subfigure}{0.49\textwidth}
        \centering
        \includegraphics[width=\textwidth]{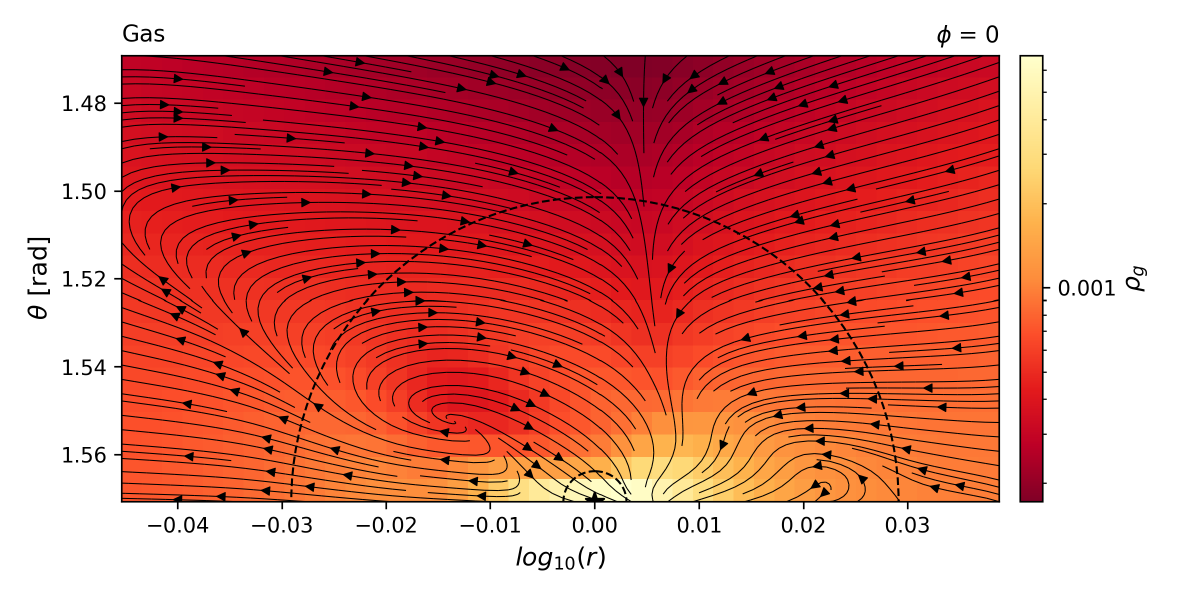}
        \caption{Gas accretion, $t=2000$ orbits}
        \label{fig: Gas acc, 2000 orbits}
    \end{subfigure}
    \begin{subfigure}{0.49\textwidth}
        \centering
        \includegraphics[width=\textwidth]{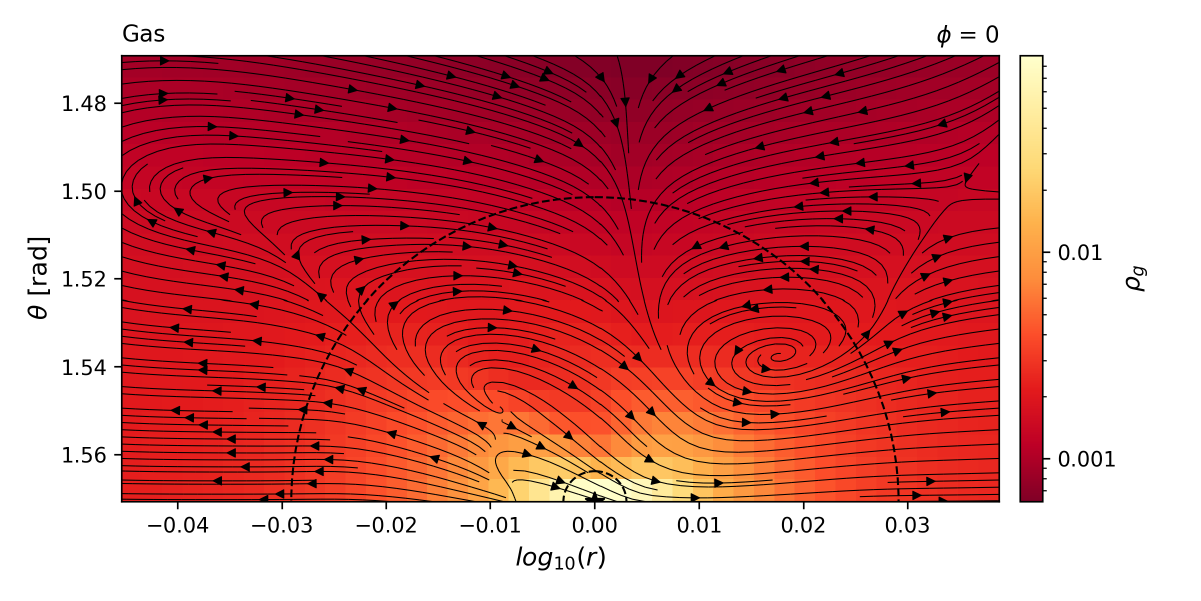}
        \caption{No gas accretion, $t=2000$ orbits}
        \label{fig: No gas acc, 2000 orbits}
    \end{subfigure}
    \caption{Streamline plots for gas in the simulations with standard parameters, $\alpha = 10^{-3}$ and $\mathrm{Sc} = 1.0$ for, left: simulations including gas accretion, right: simulations excluding gas accretion, where the top row shows results at 1000 orbits, and the bottom at 2000 orbits.    
    As in Fig.~\ref{fig: Meridional Motion}, arrows show the motion in the $(\mathrm{log}_{10}(r),\theta)$ plane at the azimuthal location of the planet, $\phi = 0.0$, where the planet is denoted by a star symbol, with two surrounding dotted circles for the sink and Hill sphere of the planet respectively. The background colormap shows the density of the gas in the slice. While the radial location further from the star (right) has signatures of the meridional motion, this feature is not seen on the other side.}
    \label{fig: Gas Meridional Flows}
\end{figure*}

Here we find that meridional motion flows in the fiducial simulation are induced by the presence of a planet across essentially all grains simulated; the exception to this is the largest $\mathrm{St} = 0.1$ dust grains, where the gap region is significantly depleted in these grains and vertical settling dominates over any radial motion. We show the meridional flow patterns for $\mathrm{St} = 10^{-2}$ and $\mathrm{St} = 10^{-3}$ grains in Fig.~\ref{fig: Meridional Motion}, and note that the profiles for the $\mathrm{St} = 10^{-4}$ and $\mathrm{St} = 10^{-5}$ are similar in morphology. 
The gas morphology for this fiducial simulation including gas accretion is also similar to the smaller dust species $\mathrm{St} = 10^{-3}$ shown in Fig.~\ref{fig: Meridional Motion} at 2000 orbits, and this motion is largely unchanged from the flow at 1000 orbits as shown in Fig.~\ref{fig: Gas Meridional Flows}. However, in the case without gas accretion included, the meridional flows show a notable evolution from 1000 orbits to 2000 orbits, moving away from the standard meridional motion, to one where the inner flow dominates such that the vertical flow does not reach as far into the planetary Hill sphere. This difference is presumably because the density in the Hill sphere reaches a steady state and stops accreting in the case without gas accretion, while it continues circulating and accreting when gas accretion is included.

While these profiles are expected for our planet and disc parameters, we can also consider how the morphology changes in simulations with varying disc turbulence, and weaker dust diffusion. We find that in both the $\alpha = 10^{-2}$ and $\alpha = 3\times10^{-3}$ simulations, signatures of the meridional circulation also manifest for the $\mathrm{St} = 0.1$ particles that are no longer significantly depleted in this region. These have the form of a full meridional morphology either side of the planet for the simulations excluding gas accretion, and a weaker meridional flow for the $\alpha = 3\times10^{-3}$ simulation with gas accretion included. This is unlike that of the $\mathrm{St} = 0.1$ grains in the $\alpha = 10^{-3}, \ \mathrm{Sc} = 1.0$ simulations where the vertical motion dominated in the case of the severely depleted gap region. The fact that this meridional circulation is apparent in all dust species run with these larger alpha parameters, suggests that the meridional motion should be present where the gap region has a non-negligible dust density, depending on the dust grain size and the turbulence of the disc for this planetary mass.

In our weaker dust diffusion simulations with $\mathrm{Sc} = 10.0$, the gas motion is unchanged from the fiducial $\mathrm{Sc} = 1.0$ case, and we continue to see the meridional circulation across the dust simulated here with $10^{-4} \leq \mathrm{St} \leq 10^{-2}$. We do, however, note that although the streamline morphology is similar to the $\mathrm{Sc} = 1.0$ simulations with and without gas accretion respectively, the dust density is more depleted within this gap region, especially at higher latitudes where $\mathrm{St} = 10^{-2}$ grains have densities close to the density floor. Running these grains for longer timescales would result in the complete depletion of this species in the vicinity of the planet and as a result this motion would no longer be seen. 

Recent 3D simulations have highlighted that material passing into the circumplanetary disc, and potentially accreting onto the planet, does so via this meridional motion, such that it will travel via the poles of the circumplanetary disc to the planet, and away from the planet in the disc mid-plane. While this was previously shown for $\mathrm{St} = 10^{-3}$ dust by \citet{Bi2021} and in $\mathrm{St} = 2\times 10^{-2}$ dust in \citet{Szulagyi2022}, we show this here over a wider range of particle sizes and under different disc parameters.

\subsection{Inner Disc Material Origins}\label{Sect 3: Inner Disc Material Origins}

Material that makes it through to the inner disc is material that will be available to potentially contribute to the formation of terrestrial planets. Hence, in order for us to understand more about their origins and any implications of the paths taken through the gap, we need to determine the origins of this material in the disc. Furthermore, investigating the trajectory of particles through the gap is important due to the impact that heating along their journey, such as from the accretion luminosity of the planet, may have on the chemical composition of these grains. 

To do this we tracked particles from a location outside the outer gap edge but inside the location of the pressure maximum forwards in time. Of particular interest are the paths of material from the outer disc whose final tracked positions lie in the inner disc. 

\subsubsection{Gas motion through the gap}\label{Sect 3: Inner Disc Material Origins - Gas Motion}

Splitting the initial heights of particles in the outer disc into windows of $0.5$H, a proportion of gas particles from all initial disc height windows in the fiducial simulation, with $\mathrm{Sc}=1.0$ and $\alpha = 10^{-3}$, are able to cross the gap. A sample of this behaviour is seen in Fig.~\ref{fig:gas (rcostheta,rsintheta) plot} for gas particles originating from $1.5-2.0$H that make it through the gap region. 
\begin{figure}
	\includegraphics[width=\columnwidth]{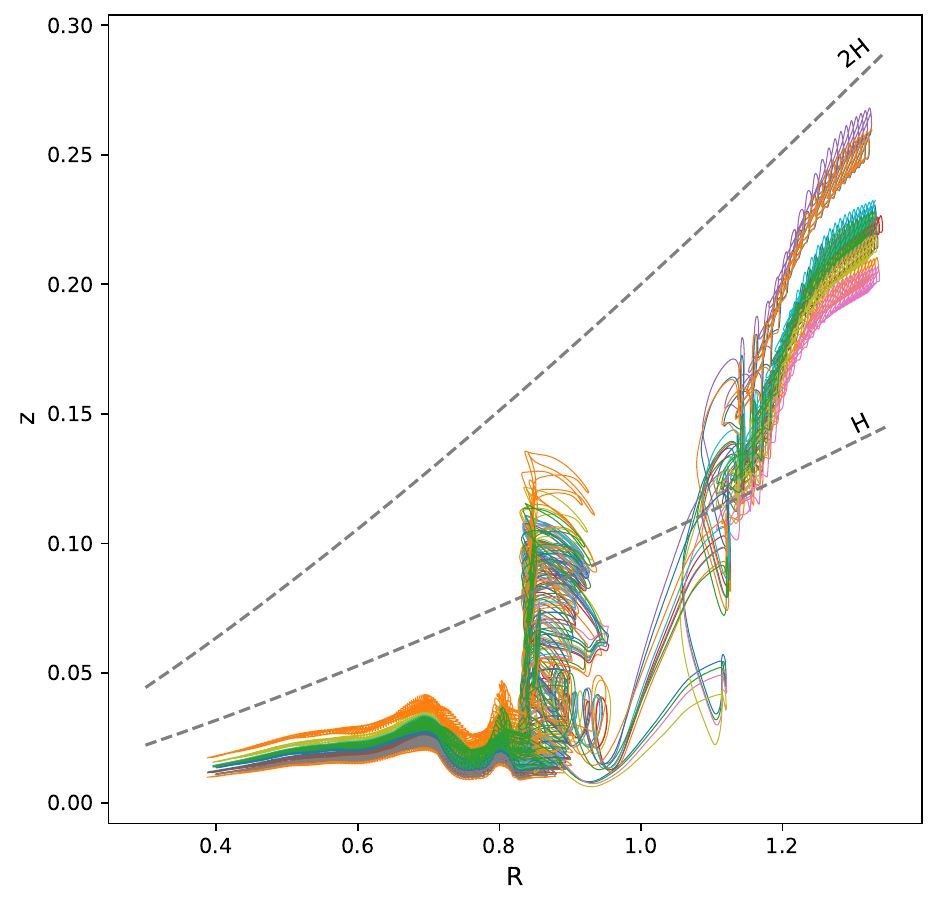}
    \caption{Gas particle tracks for material that is able to make it into the inner disc in the fiducial simulation including gas accretion for particles starting with initial heights between $1.5-2.0\mathrm{H}$. Each line, and also colour, shows a different particle path, with the general trend being that particles drop a scale height towards the gap region before crossing the gap, falling in altitude. The dotted lines indicate the $1\mathrm{H}$, and $2\mathrm{H}$ scale heights of the disc. The paths are tracked to a radial value minimum of $r = 0.4$ so as not to include motion within the Stockholm boundaries.}
    \label{fig:gas (rcostheta,rsintheta) plot}
\end{figure}
Examining this behaviour more closely, this material ends up close to the mid-plane in the inner disc, with final vertical positions between $0.0-1.0$H for all gas particle paths that make it through. This result implies that the inner disc gas is composed of a mix of gas originating in the outer disc, and the gas that was already available in this region prior to planetary gap formation. 

\begin{figure}
\begin{subfigure}{\columnwidth}
  \centering
  \includegraphics[width=\columnwidth]{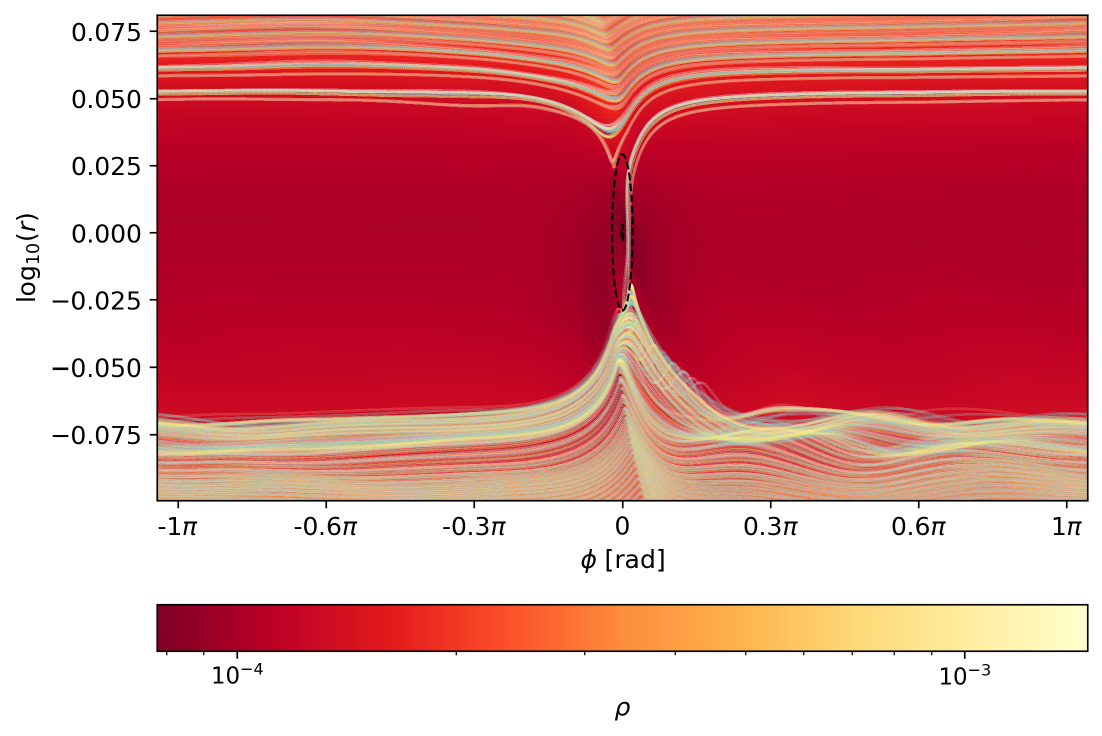}
  \caption{Gas}
  \label{fig: Gas Horseshoes}
\end{subfigure}
\begin{subfigure}{\columnwidth}
  \centering
  \includegraphics[width=\columnwidth]{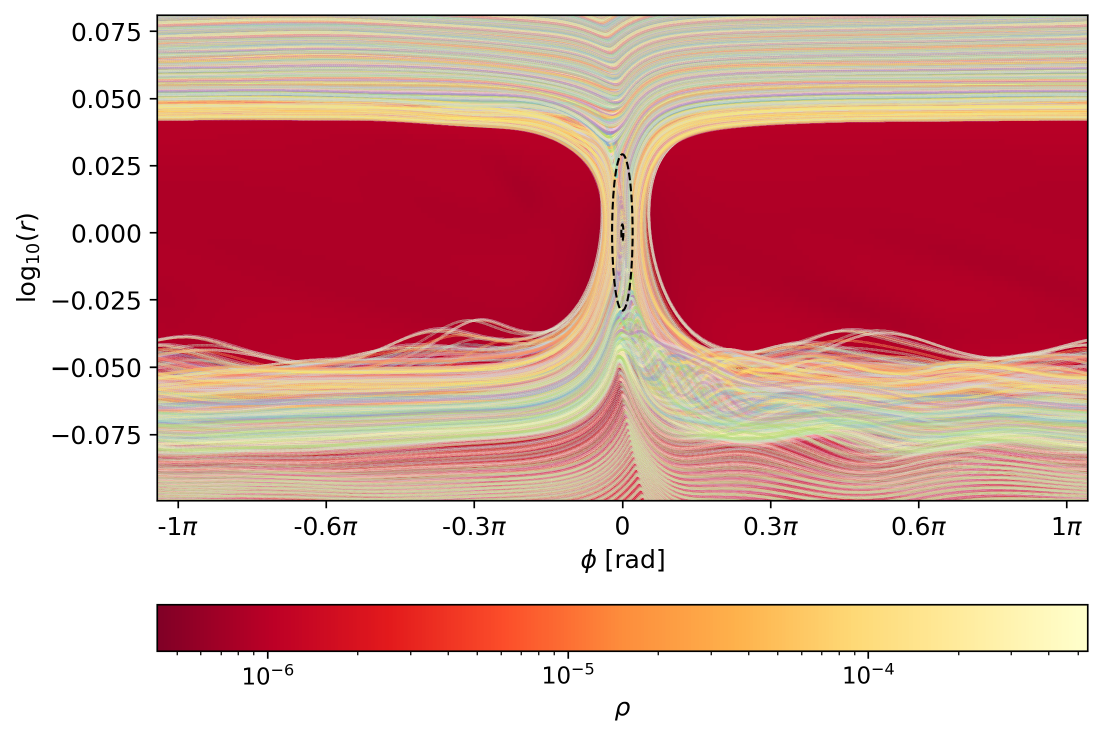}
  \caption{$\mathrm{St} = 10^{-5}$}
  \label{fig: St10-5 Horseshoes}
\end{subfigure} 
\caption{Particle paths of material that successfully makes it from the outer disc into the inner disc region, for gas (top) and $\mathrm{St} = 10^{-5}$ dust (bottom). Each line shows the path taken by a particle passing through the gap region, displayed over the density of the same species at $2.0$H in the disc, with all particles shown originating from initial heights between $1.5-2.0$H. Ellipses depicting the planetary Hill sphere and sink radius are shown above the planetary location at $\phi = 0, r = 1$. The y-axis is sliced to zoom in on these paths through the gap region, therefore this does not span the entire radial domain.}
\label{fig:paths through via horseshoe region}
\end{figure}

Regardless of initial height in the outer disc, all gas particles that traverse the gap do so via horseshoe orbits, an example of which is shown in Fig.~\ref{fig: Gas Horseshoes}.
In addition, all particles that make it through the gap region with initial heights below $2.0\mathrm{H}$ do so passing through the Hill sphere of the planet, as seen in Fig. \ref{fig: Fig 7, gas}. For higher initial heights, this proportion of trajectories passing through the Hill sphere of the planet decreases, with cumulative frequency values of $0.79$ for gas with initial heights $2.0\mathrm{H} < \mathrm{H}_{i} \leq 2.5\mathrm{H}$, and $0.22$ for initial heights $2.5\mathrm{H} < \mathrm{H}_{i} \leq 3.0\mathrm{H}$.
\begin{figure*}
\begin{subfigure}{.36\textwidth}
  \centering
  \includegraphics[width=.9\linewidth]{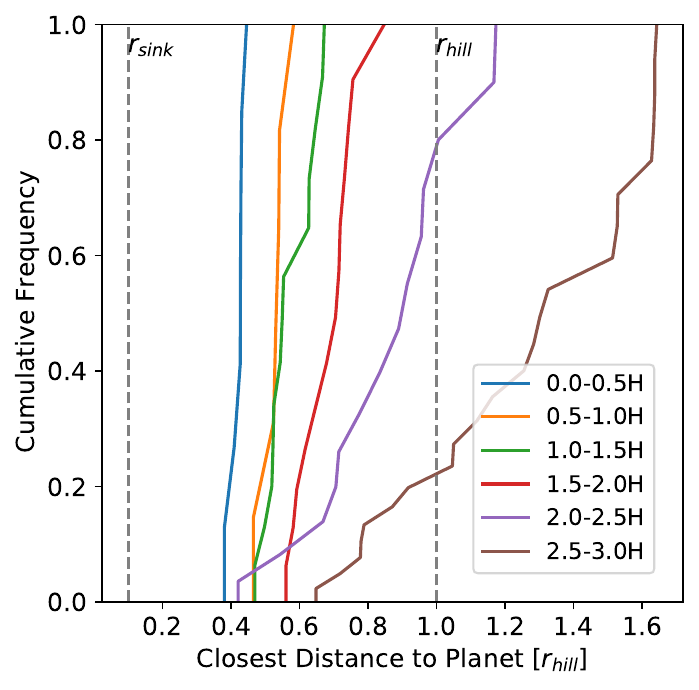}
  \caption{Gas}
  \label{fig: Fig 7, gas}
\end{subfigure}
\hspace{-0.85cm}
\begin{subfigure}{.36\textwidth}
  \centering
  \includegraphics[width=.9\linewidth]{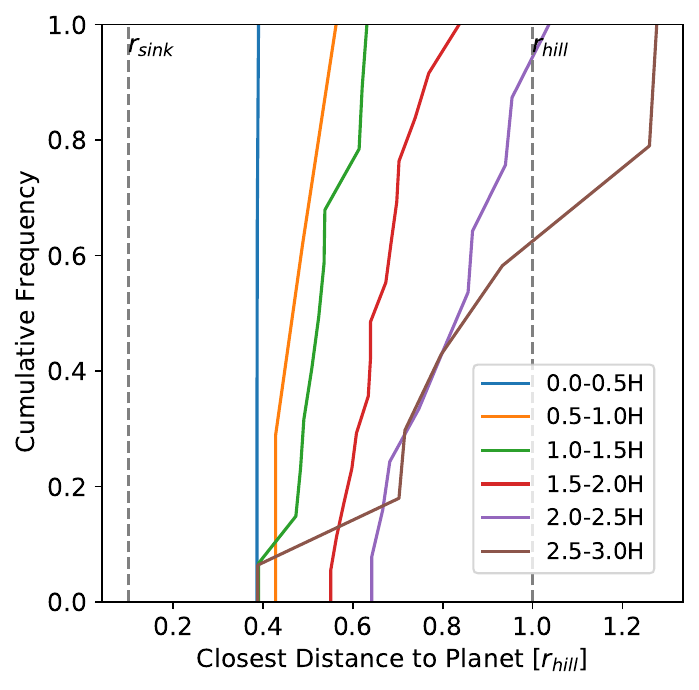}
  \caption{$\mathrm{St} = 10^{-5}$}
  \label{fig: Fig 7, 10-5}
\end{subfigure} 
\hspace{-0.85cm}
\begin{subfigure}{.36\textwidth}
  \centering
  \includegraphics[width=.9\linewidth]{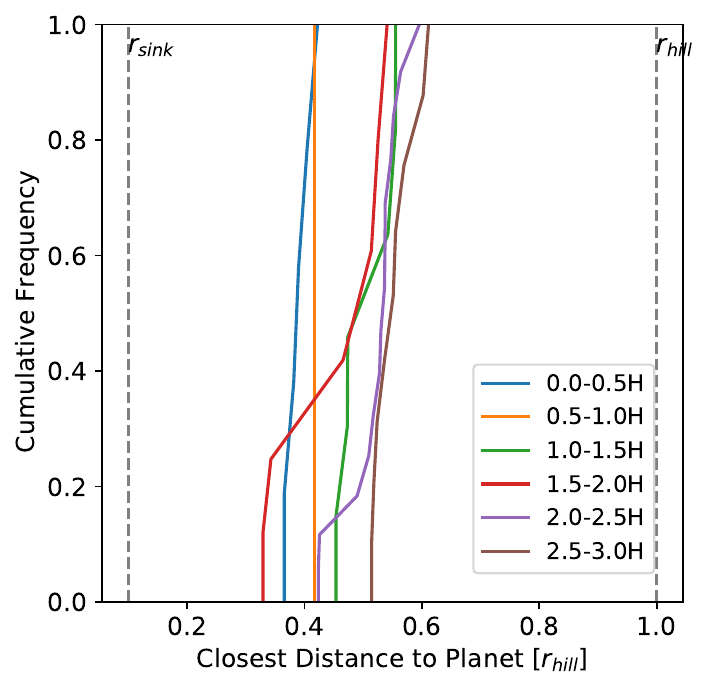}
  \caption{$\mathrm{St} = 10^{-4}$}
  \label{fig: Fig 7, 10-4}
\end{subfigure}
\caption{Plots showing the cumulative frequency of closest distances particles have to the planet for those particles that pass through the gap region. For each plot, the particles are split and labelled with their respective initial height windows from the outer disc. The vertical dashed lines depict the planet sink radius, $r_\mathrm{sink}$, and Hill sphere, $r_\mathrm{hill}$, respectively. The leftmost plot (a) shows this for the gas in the fiducial simulation (including gas accretion) with $\alpha = 10^{-3}$ and $\mathrm{Sc} = 1.0$. (b) shows this for the smallest dust $\mathrm{St} = 10^{-5}$ in the same simulation, and (c) shows this for the $\mathrm{St} = 10^{-4}$ dust.}
\label{fig: Cumulative frequencies of closest distances to planet}
\end{figure*}
 
\begin{figure*}
\begin{subfigure}{.49\textwidth}
  \centering
  \includegraphics[width=8.5cm, height=7.5cm]{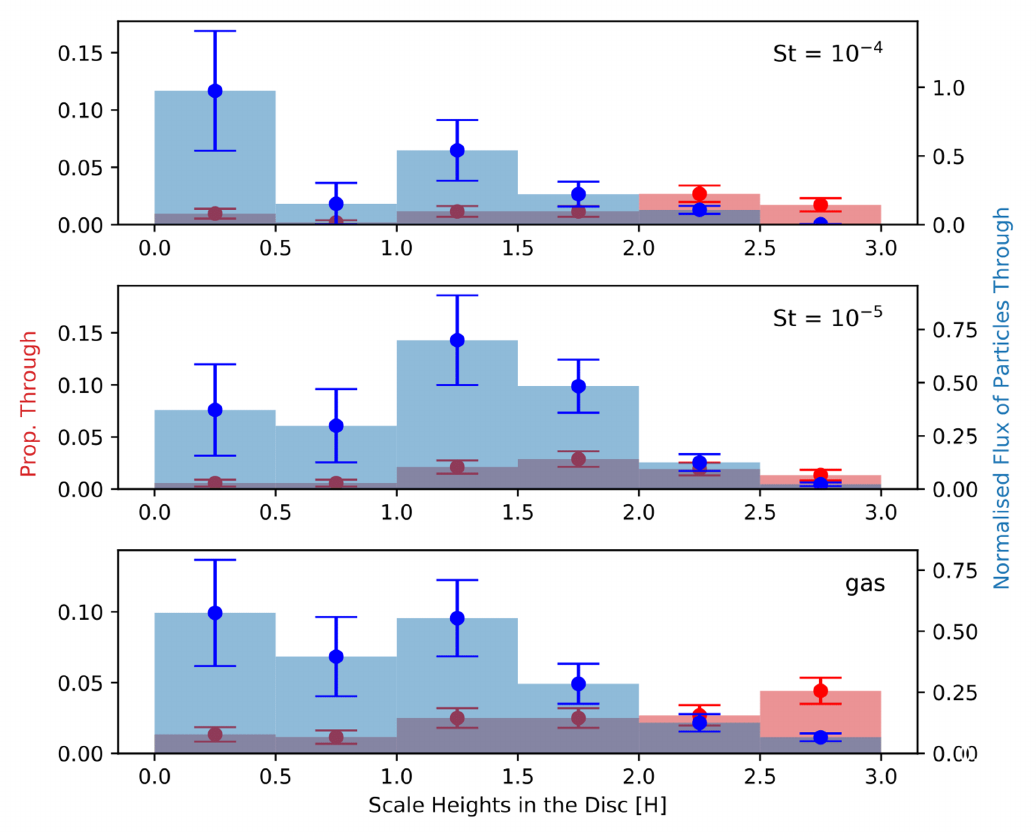}
  \caption{Gas Accretion}
  \label{fig: Gas Acc Hist}
\end{subfigure}
\begin{subfigure}{.49\textwidth}
  \centering
  \includegraphics[width=8.5cm, height=7.5cm]{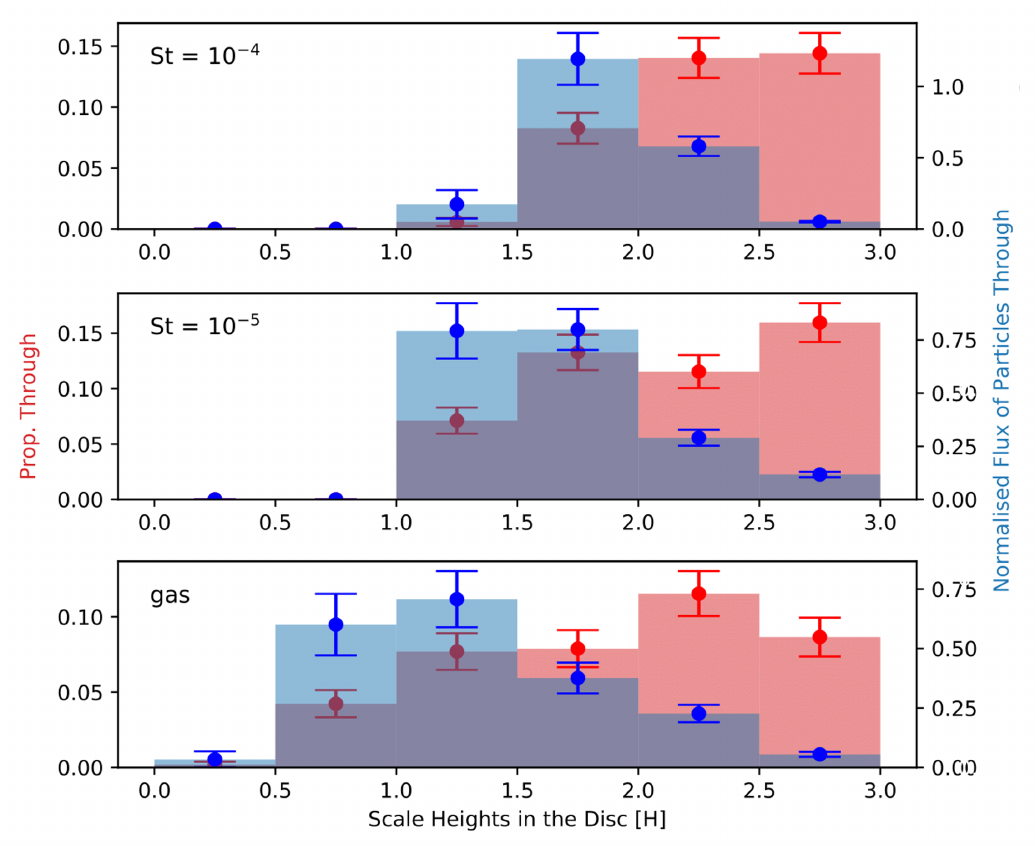}
  \caption{No Gas Accretion}
  \label{fig: No Gas Acc Hist}
\end{subfigure}
\caption{Red histogram: histogram of fraction of particles in each bin of initial half scale height that make it through the gap. Blue histogram: histogram of fraction of total flux of particles that make it through the gap that originate in each bin of initial height, obtained by weighting by the density average in each initial scale height bin and re-normalising to unity. The results for the fiducial simulation with gas accretion are shown on the left, and that of the simulation excluding gas accretion on the right, where the dust of $\mathrm{St} = 10^{-4}$ is shown on top, $\mathrm{St} = 10^{-5}$ in the middle, and gas results on the bottom in each case.}
\label{fig: Counts and Flux}
\end{figure*}

The proportion of gas particles from this initial radius outside the planet that make it through the gap region generally increases with distance above the mid-plane. From this we can determine where most of the material that makes it through the gap comes from by multiplying the probability of success with the half scale height window averaged density at the initial radial location in the outer disc. From this we find that the greatest amount of gas passing through the gap originates from scale heights $\mathrm{H}_{i} \leq 1.5\mathrm{H}$, as seen in Fig.~\ref{fig: Gas Acc Hist}.

\subsubsection{Dust motion through the gap}\label{Sect 3: Inner Disc Material Origins - Dust Motion}

As expected when investigating the dust species that make it through the gap region, the small grains have behaviour similar to the gas, being much more efficient in passing through the gap than large dust, as previously seen when examining the dust filtration in Section \ref{Sect 3: Dust Filtration}. As with the gas, the small grains, with $\mathrm{St} = 10^{-4}$ and $\mathrm{St} = 10^{-5}$, that make it through the gap do so following the horseshoe paths in agreement with \citet{Binkert2023}. These do so largely passing through the Hill sphere of the planet when considering the face down view of the disc, an example of which is shown in Fig.~\ref{fig: St10-5 Horseshoes} for the $\mathrm{St} = 10^{-5}$ dust. 

It can be seen in Fig.~\ref{fig: Cumulative frequencies of closest distances to planet} that most particles with $\mathrm{St} \leq 10^{-5}$ that move through the gap region pass within this planetary Hill radius. Here, the particles starting from higher scale heights in the disc do not pass through the gap as closely to the planet as those originating from lower scale heights. Particles that pass through the gap with initial heights $\mathrm{H}_{i} \leq 2.0\mathrm{H}$ all pass through the Hill sphere of the planet, with 94\% of particles from initial heights $2.0\mathrm{H} < \mathrm{H}_{i} \leq 2.5\mathrm{H}$ and 62\% of particles from  initial heights $2.5\mathrm{H} < \mathrm{H}_{i} \leq 3.0\mathrm{H}$ passing through the Hill sphere. Note that this is a larger proportion through the Hill sphere for these higher initial scale heights than for the gas above. Furthermore, for the larger $\mathrm{St} = 10^{-4}$ dust, all particles that make it through the gap region pass through the Hill sphere of the planet.

Considering the proportions of this dust through the gap region referring back to Fig.~\ref{fig: Gas Acc Hist} we can see that the flux through the gap as a function of original height is broadly similar for these dust species as for the gas. The proportion through the gap region computed using our tracking method shows a tendency for reduced transport in trajectories originating near the mid-plane as the Stokes number is increased. Multiplying by the average density in the half scale height window, however, returns the highest mass flux through the gap region for the $\mathrm{St} = 10^{-4}$ grains for $\mathrm{H}_{i} \leq 0.5\mathrm{H}$, and for $\mathrm{St} = 10^{-5}$ grains between $1.0\mathrm{H} < \mathrm{H}_{i} \leq 1.5\mathrm{H}$, with mass flux through the gap for both species being markedly reduced for initial heights $\mathrm{H}_{i} \geq 2.0\mathrm{H}$.

Conversely, for the same simulation without gas accretion, the small dust grains $\mathrm{St} = 10^{-5}$ and $10^{-4}$ cannot make it through the gap region from initial heights in the outer disc of $\mathrm{H}_{i} \leq 1.0 \mathrm{H}$. Subsequently multiplying by the average density in the half scale height window shifts the highest mass flux through the gap for these particles to higher initial scale heights between $1.5\mathrm{H} < \mathrm{H}_{i} \leq 2.0\mathrm{H}$ for the $\mathrm{St} = 10^{-4}$ grains, and between $1.0\mathrm{H} < \mathrm{H}_{i} \leq 2.0\mathrm{H}$ for $\mathrm{St} = 10^{-5}$ grains as shown in Fig.~\ref{fig: No Gas Acc Hist}.

In both cases, the transport of large dust across the gap is strongly inhibited. Tracking paths of $10^{-3} \leq \mathrm{St} \leq 0.1$ dust from the outer disc inwards returns no particles across the range of initial positions run that successfully make it into the inner disc. Instead these get stuck beyond the planet in the pressure maximum region.

\subsection{Dust Diffusion vs Disc Viscosity}\label{Sect 3: Diffusion vs Viscosity}

To consider whether the dust diffusion or viscosity has a greater impact on the dust filtering through the gap region formed in the presence of an embedded planet, we explore the behaviour varying these parameters.

Referring back to Fig.~\ref{fig: Pressure Max Filtration} and Fig.~\ref{fig: Outer Radial Flux}, it is clear that reducing the dust diffusion, through the use of a larger Schmidt parameter of $\mathrm{Sc} = 10.0$, results in a dust filtering effect that is more prominent than in the fiducial case, with dust being easily prevented from making it into the gap region. Consequently these simulations have a deep gap region, with less dust flux reaching the planet and a decreased proportion of this dust accreted in Fig.~\ref{fig: Planet Filtering Fraction}.

Varying the viscosity parameters from the fiducial $\alpha = 10^{-3}$, again referring to Fig.~\ref{fig: Pressure Max Filtration} and Fig.~\ref{fig: Outer Radial Flux}, show that a smaller alpha viscosity parameter leads to a greater filtration of large dust grains into the gap region, with the fiducial value of alpha resulting in the greatest prevention of transport though the gap region for the largest grains with initial $\mathrm{St} = 0.1$. At the largest alpha parameter value considered here of $\alpha = 10^{-2}$, the large dust has behaviour against the trend in other simulations, with the largest dust being able to pass into the gap region entirely as the viscosity acts to fill in the gap region. This behaviour is expected as the lack of a pressure maximum formed allows the large dust to move inwards without experiencing trapping beyond the planet, where deeper gaps result in greater filtration of larger dust grains. The impact on the small Stokes numbers is less significant, with values of the $\mathrm{St} = 10^{-5}$ dust prevented getting through the gap region due to accretion onto the planet converging to $0.82$, $0.76$ and $0.59$, for the alpha parameters (without gas accretion), $\alpha = 10^{-3}, \ 3\times 10^{-3}$ and $10^{-2}$ respectively. This can be explained by the time required for particles to cross the accretion region of the planet since the radial speed of tightly coupled grains has a linear dependence on the disc viscosity. Therefore the grains in the $\alpha = 10^{-2}$ case cross the accretion cross section of the planet on the shortest timescales, leading to the least time spent in this region and less accretion onto the planet as a result.

\subsection{Gas Accretion vs No Gas Accretion}\label{Sect 3: Comparisons to Gas Accretion Simulations}

Our standard simulation includes gas accretion onto the planet, however, since gas accretion changes the gap structure in these discs, we include a set of simulations without gas accretion to directly examine the influence that gas accretion has on the transport of material through the disc. 

Excluding gas accretion both increases the surface density of the gas in the gap region and the pressure support at the location of the planet, resulting in a difference in the transport through the disc for both the gas and dust. We find the proportion of particles crossing the gap to be much higher than in the simulation including gas accretion onto the planet, as shown in Fig.~\ref{fig: Counts and Flux}. This is likely due to stronger gas flows from the outer disc to the planet's radial location in this case. We also find that excluding gas accretion, and therefore increasing the pressure support at the planet location, prevents particles originating from lower initial scale heights in the disc from passing through the gap region, shifting the highest mass flux through the gap region away from the mid-plane when accounting for the densities in the outer disc. This is demonstrated in Fig.~\ref{fig: No gas acc, 2000 orbits} where flow into the planetary Hill sphere from the side further from the star enters from higher scale heights. This can be compared to the simulation with gas accretion (see Fig.~\ref{fig: Gas acc, 2000 orbits}) where the gas flow is similar to the flow patterns shown for small dust particles in Fig.~\ref{fig: Meridional Motion} with motion into the Hill sphere from closer to the mid-plane. As a result, excluding gas accretion shifts the flow of gas that enters the planetary Hill sphere and through the gap region to higher scale heights, and likewise this expands to the small dust grains that are well-coupled to the gas.

\section{Discussion}\label{Sect 4: Discussion}

Here we discuss the implications of our results on the potential chemistry of the inner disc region. In what follows it should be kept in mind that the simulations presented here have been conducted assuming a locally isothermal disc and the exclusion of a more detailed temperature structure and additional physics may change the dust filtering and motion through the gap region.

\subsection{Implications of Grain Transport for the Chemistry of the Inner Disc}\label{Sect 4: Origin Implications for Chemistry}

To understand the impact of this movement of grains across the region of the planet in terms of the composition of the disc we need to consider what gets through the gap region and in what quantities. 

Dust particles with $\mathrm{St} < \alpha$ have been shown to make it through the gap in 2D, and most recently in 3D as \citet{Karlin2023} showed that large dust grains have inefficient accretion into a circumplanetary disc due to dust filtration at the pressure maximum at the outer edge of the gap. Here our simulations are run for longer timescales and we investigate the dominant mechanism behind this transport through the gap region by varying our disc diffusion and disc viscosity parameters separately, as seen in Section \ref{Sect 3: Diffusion vs Viscosity}.

Whereas the simulations which do not account for gas accretion onto the planet show that the material that flows through the gap preferentially originates from the  upper layers of the disc, once planetary gas accretion is included this tendency is weakened (Fig.~\ref{fig: Counts and Flux}). Indeed, the largest particles that flow through the gap (see Fig.~\ref{fig: Outer Radial Flux}), which can be expected to contain the majority of the dust mass flowing through the gap, predominantly originate from less than $0.5\mathrm{H}$. Given that the disc aspect ratio used in the simulations corresponds to placing the planet at around $100$ AU, such grains originate from regions which are well below the CO vertical snowline \citep{Dutrey2017, Pinte2018} and it can thus be expected that these grains arrive in the planet induced gap region with icy CO mantles. 

With the influence of dust filtering on large grains, we should expect a reduced dust-to-gas ratio of material accreting onto the planet as a result, and would hence have lower metallicity material accreting on the planet and moving to the inner disc regions. However, \citet{Szulagyi2022} find that mm-grains can get trapped more easily in the CPD than the gas suggesting that this could lead to an enrichment in solids within this region compared to the circumstellar disc. Due to resolution limitations in both their work and ours, monitoring the CPD mass itself will not produce correct results. Instead they monitor the change in the Hill sphere mass over a planetary orbit. Doing the same here over the last 15 planetary orbits, we find that our smallest dust grains have lower Hill sphere masses than the gas mass divided by the initial gas-to-dust ratio of 100. Larger dust grains are increasingly depleted in this region, and there are no significant changes over the time period monitored, a result of reaching a near-steady state. Consequently the Hill sphere appears to be less enriched in solids than the outer disc region as a result of the dust filtering, the difference of which may be due to the longer run times used here that are an order of magnitude longer.

\subsection{Accretion Luminosity Impact on Chemistry}\label{Sect 4: Accretion Impact on Chemistry}

It is important to investigate how the dust crosses the gap to understand the potential consequences this has on the chemical composition of this material, considering the processes that this material may be subject to on its journey. For example, additional thermal processing from the accretion luminosity of the planet can lead to sublimation of volatiles originally stored in ice on dust particles en route to the inner disc. 

We have previously (\ref{Sect 4: Origin Implications for Chemistry}) argued that the bulk of grains arriving in the gap region for the planet modelled (corresponding to a Jupiter mass planet at $\sim 100$ AU) would be coated with icy CO mantles. \citet{Cleeves2015} explored the thermal influence of such a planet in the case of an accretion rate onto the planet of $\dot{M}_p = 10^{-8} M_\odot$ year$^{-1}$ which is similar to that of our fiducial simulation of $\dot{M}_p \sim 2 \times 10^{-8} M_\star$ year$^{-1}$. \citet{Cleeves2015} showed that the temperature in the gap region is around $50\%$ larger than that of the ambient mid-plane disc and that moreover, within the planetary Hill sphere the temperature is enhanced by a further $\sim30\%$. When this is combined with the result from our simulation that the majority of paths crossing the planet-carved gap have passed through the Hill sphere of the planet, it is inescapable that CO will come off the grains at this point. The ultimate fate of CO vapour being released within the planetary Hill sphere cannot however be addressed by our simulations since they do not track diffusive motions in the gas. Future implementation of live volatiles within our simulations would be needed to determine whether the released CO is predominantly accreted onto the planet or conveyed to the inner disc where it would re-condense on grains.

\subsection{Simulation Caveats}\label{Sect 4: Simulation Caveats}

Work has been done to highlight the importance of including radiative transfer in these models. \citet{Szulagyi2022} include radiative transfer in their simulations and find that the gap width varies at different heights due to the vertical temperature gradient. They comment that this effect also influences the dust distribution at different heights in the disc, and therefore without taking into account all heating and cooling mechanisms here as they conducted, the dust filtering may be subject to change. This is supported by work done iterating between hydrodynamic and radiative transfer simulations by \citet{Chen2024}, finding that deeper and wider gaps are found compared to the case without iterative radiative transfer. \citet{Ziampras2023} further stress the importance of realistic cooling in their simulations, showing that radiative cooling affects the structure of the gaps formed. We note that \citet{Cummins2022} show with sub-isolation mass planets that the presence of vortices may make a difference to the dust accretion, however, we would need to include the luminosity from the planet to determine if they are present here.

The exclusion of dust coagulation and fragmentation is also likely to impact the total amount of material that makes it through the gap. \citet{Stammler2023} suggest that dust fragments in the outer disc beyond the gap region, and find that in fragmenting, this smaller dust is then able to pass through the gap region into the inner disc. They find that even in the presence of a Jupiter mass planet, the entire dust mass reservoir from the outer disc is able to filter through into the inner disc in their 1D models, with larger planet masses inducing deeper gaps that subsequently require longer times for this process to occur. While we don't include dust growth and fragmentation within our simulations, the result \citet{Stammler2023} find would suggest that eventually all the dust that appears to be stuck in the outer disc in our simulations would fragment to smaller sizes that are able to pass through the gap region, depleting the outer disc. Providing this fragmentation occurs in the 3D case, our simulations agree that the outer dust disc will be depleted through the transport of small grains through the gap region, albeit with a large proportion of this small dust being accreted by the planet. Since the larger dust grains will fragment, adding to the mass of the smaller grain populations, \citet{Stammler2023} also suggest that the presence of a giant planet is therefore not sufficient to explain the dichotomy of non-carbonaceous and carbonaceous chondrites in our solar system, and that there would need to be other contributing factors.

Finally, our planet is kept on a fixed, circular orbit with no planetary migration included. For massive planet masses we expect Type II planet migration to play a part in their orbital evolution. \citet{Meru2019} found that planet migration changes the gap structure compared to a non-migrating planet. Their study was conducted for sub-isolation mass planets, however, in a similar way, varied gap structure may be seen for more massive planets. Consequently, one might expect different results including this migration.

\section{Conclusions}\label{Sect 5: Conclusion}

In this paper we have studied the impact of a massive embedded planet in a protoplanetary disc on the surrounding material using 3D global hydrodynamic simulations conducted with \textsc{FARGO3D} including gas and multiple dust species. We included a planet with a mass of twice the pebble isolation mass for the aspect ratio run of $h_0 = 0.1$, such that the planet mass is $M_p = 0.001 M_\star$. A range of fixed dust grain sizes in each simulation are run, and we refer to these by their initial Stokes numbers set at the planet location at the start of the simulations. We include dust accretion onto the planet in all simulations, with a set including gas accretion and a set excluding gas accretion to compare the impact this has on material evolution in the disc.

We find meridional motions in the planet vicinity are present for all gas and dust species providing that the gap region is not completely depleted. Furthermore, we investigated the impact of varying the dust diffusivity and alpha viscosity on the dust filtering. Reducing the dust diffusivity ($\mathrm{Sc} = 10.0$) significantly increased the amount of dust trapped beyond the planetary gap while increasing the alpha parameter allows larger dust grains to flow through the gap in agreement with simple expectations based on comparing $\mathrm{St}$ and $\alpha$.

Once the simulations had evolved to a near-steady state we conducted particle tracking to follow the average motion of both gas and dust. We find that for the fiducial simulation with $\mathrm{Sc} = 1.0$ and $\alpha = 10^{-3}$, small grains and gas are able to traverse through the gap region from all initial heights in the outer disc with maximal flux of material between $0.0 - 1.5\mathrm{H}$ for the case with gas accretion, to higher initial scale heights between $1.0 - 2.0\mathrm{H}$ without gas accretion. Considering the closest distances particles get to the planet on their journey through the gap, these particles do so largely passing through the planetary Hill sphere. Therefore, large proportions of this material flowing through the gap region would experience heating due to the accretion luminosity incident from the planet on their journey through the disc. On the other hand, the larger dust grains with $\mathrm{St} \geq 10^{-3}$ become stuck beyond the planet location for all initial heights.

This key result that small grains can still make it through a planetary-induced gap, and do so largely by passing through the Hill sphere of the planet has consequences for the composition of material that is able to both accrete onto the planet itself, and the composition of material that is able to make it into the inner disc for further planetary growth. We conclude that material available in this region for additional planetary growth should be formed of a mixture of material present in the inner disc during the gap formation, and material that has been able to permeate through the gap region from the outer disc, composed of smaller dust grains. Since larger dust grains do not make it into the inner disc we have a reduced dust-to-gas ratio of material making it into the gap, accreting on the planet and also in the inner disc. However, since these simulations are run without including dust growth and fragmentation, it may be the case that large dust grains trapped beyond the planet fragment, and subsequently a larger proportion of the total dust could pass through as seen in the 1D simulations run by \citet{Stammler2023}. The addition of dust coagulation and fragmentation will be necessary to investigate this. 

Finally, we note that our simulations with $H/R = 0.1$ correspond to a planet at $\sim 100$ AU, which allows us to comment on the influence of the planet on the disc thermochemistry. At this radius, the planet is located outside the radial CO snowline; moreover, we have shown that the upper size range of the grains that can pass through the gap originate in the cold mid-plane regions of the outer disc, below the vertical CO snow-line. On the other hand, the significant (factor two) boost in temperature expected within the Hill sphere of the planet, combined with the fact that the flow trajectories are concentrated in the Hill region, implies that CO is very likely to be evaporated from the grains as they pass through the Hill region. This injection of large quantities of CO vapour in the Hill region is likely to be a key factor in setting the C/O ratio of the nascent planet and the inner disc. However, further simulations, employing tracer particles for volatile species, are required in order to assess whether the CO released in this region is accreted by the planet or instead recondenses in the disc interior to the planetary orbit.

\section*{Acknowledgements}

HJP thanks the Science and Technology Facilities Council (STFC) for a Ph.D. studentship and also thanks Oliver Shorttle for fruitful discussions. RAB was supported by a Royal Society University Research Fellowship. CJC has been supported by the Science and Technology Facilities Council (STFC) via the consolidated grant ST/W000997/1 and also by the European Union’s Horizon 2020 research and innovation programme under the Marie Sklodowska-Curie grant agreement No. 823823 (RISE DUSTBUSTERS project).

This work used the DiRAC Data Intensive service (CSD3 [*]) at the University of Cambridge, managed by the University of Cambridge University Information Services on behalf of the STFC DiRAC HPC Facility (www.dirac.ac.uk). The DiRAC component of CSD3 at Cambridge was funded by BEIS, UKRI and STFC capital funding and STFC operations grants. DiRAC is part of the UKRI Digital Research Infrastructure.

\section*{Data Availability}

The code used to perform these simulations, FARGO3D \citep{Benitez-Llambay2016}, is available for download at \href{http://fargo.in2p3.fr/}{http://fargo.in2p3.fr/}.  
 



\bibliographystyle{mnras}
\bibliography{paper} 




\appendix

\section{Dust Accretion Fraction}\label{Appendix: Dust Accretion Fraction}

Here we describe the method and result determined from an investigation into the appropriate value to use for the fraction of material accreted per unit time, $f/\tau$ for the dust. As mentioned in Section~\ref{Sect 2: Accretion Prescription}, we use a fixed value for the gas of $f/\tau = \frac{5}{\pi}\Omega_p$, noting that \citet{Li2021} found accretion rate convergence in the gas for different $f/\tau$ values providing the sink radius $ \Delta \leq 0.1r_H$. An investigation for the dust, however, has not previously been conducted.

For this investigation, we used 2D simulations, and varied the $f/\tau$ factor, for planet masses in the sub-isolation, and super-isolation mass regimes, with the equivalent of an $5M_{\earth}$, and $1M_J$ for a central star of solar mass. In these simulations, the aspect ratio of the discs are $h=0.056$, and $h=0.05$, respectively. To determine the most appropriate value, we varied the factor $f/\tau$ and examined the accretion rate onto the planet for several dust species once the system has reached a quasi-steady state. 

\begin{figure*}
    \centering
    \includegraphics[width=\textwidth]{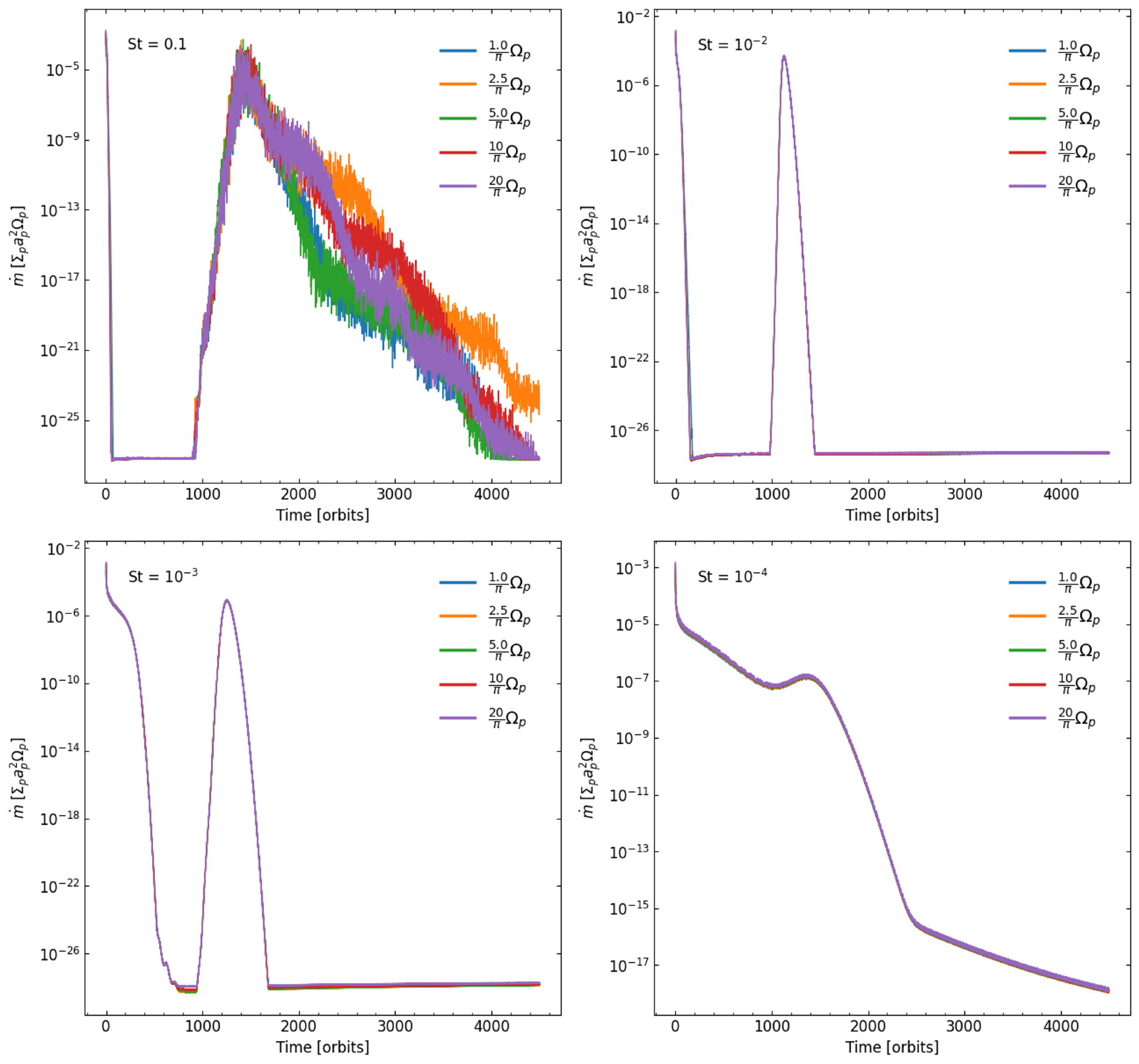}
    \caption{Dust accretion rates, $\dot{m}$, as a function of time in the Jupiter mass simulation, with dust of $\mathrm{St} = 0.1$ (top left), $\mathrm{St} = 10^{-2}$ (top right), $\mathrm{St} = 10^{-3}$ (bottom left), and $\mathrm{St} = 10^{-4}$ (bottom right). For each dust species, several lines are plotted for the accretion rates with varying $f/\tau$ values, from $\frac{1.0}{\pi} \Omega_p \leq f/\tau \leq \frac{20}{\pi} \Omega_p$, as denoted in the top right corners. For all but the largest dust species, with $\mathrm{St} = 0.1$, the dust accretion rates are unaffected by the varying the dust accretion factors, and in this largest dust the general trends remain similar.    
    }
    \label{fig: Jupiter Mass Dust Factor Convergence}
\end{figure*}

Running this for the $1M_J$ planet mass, we examined values either side of the gas value above, testing values for the dust of $f/\tau = \frac{1.0}{\pi}\Omega_p$ , $\frac{2.5}{\pi}\Omega_p$, $\frac{5.0}{\pi}\Omega_p$, $\frac{10.0}{\pi}\Omega_p$, and $\frac{20.0}{\pi}\Omega_p$. Fig.~\ref{fig: Jupiter Mass Dust Factor Convergence} shows the accretion rates as a function of time for these different factors across four dust species with Stokes numbers $\mathrm{St} = 0.1, 10^{-2}, 10^{-3}, 10^{-4}$. Only the largest dust shows small variations between the accretion rates, but we consider this to be acceptable given that we use the same value of the accretion factor, $f/\tau$, throughout and we are interested in understanding differences in this rate with varying planet and disc parameters. The other dust sizes show close agreement at all times and thus we could choose any of the values of $f/\tau$ for simulations in this regime. 

The apparent blip in the accretion rates starting around 1000 orbits and seen in all dust species here is likely a consequence of the lack of dust diffusion in these simulations, where this would usually clear out the horseshoe region. Instead, after the initial clearing of the gap due to accretion and planet-disc interactions, the Stokes number of the dust in this region increases due to the depletion of gas, leading to high Stokes numbers. In this regime the dust is poorly approximated as a fluid and consequently scatters within the region causing an increase in accretion as this material moves within the accretion sink. This blip in the accretion rate then slowly decreases for all species as the material is accreted. 

\begin{figure}
    \centering
    \includegraphics[width=\columnwidth]{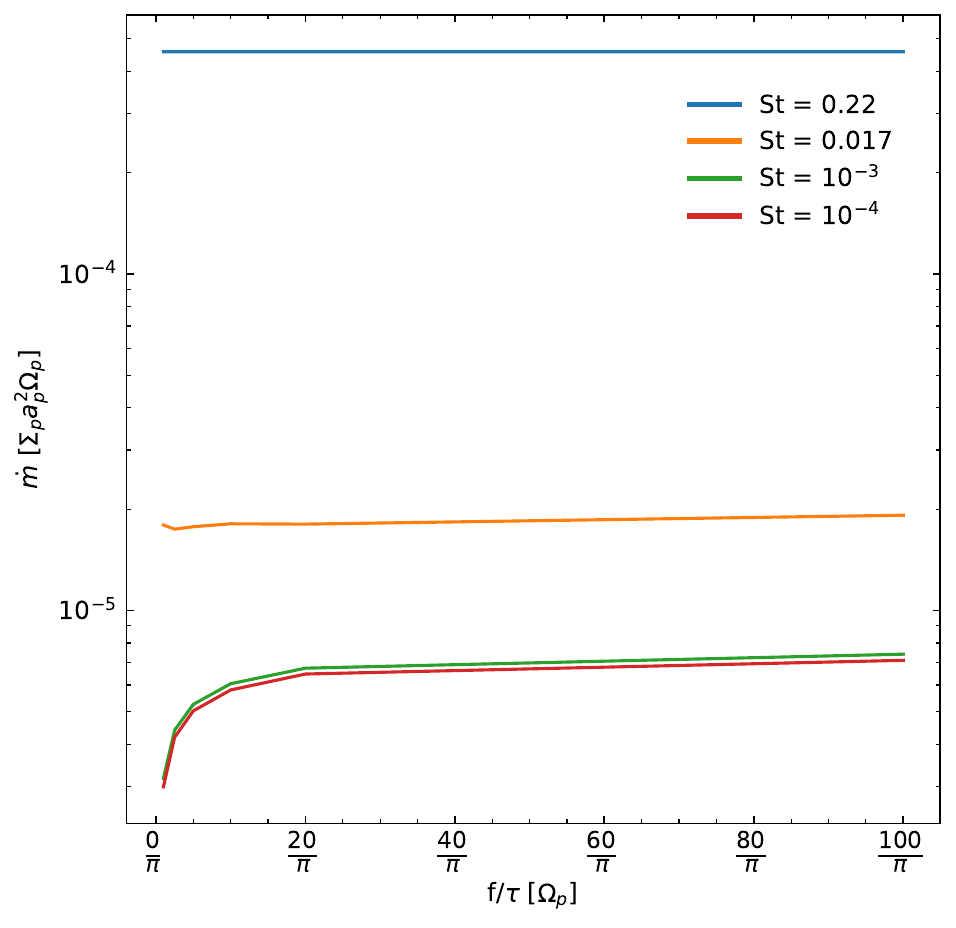}
    \caption{Dust accretion rates for dust with initial Stokes numbers, $\mathrm{St} = 0.22, 0.017, 10^{-3}$, and $10^{-4}$, as a function of accretion factor taking an average of the last 300 orbits of each simulation. We see negligible difference in the accretion rate for the larger particles, while the two smaller species converge towards large accretion factors.}
    \label{fig: 5ME Dust Factor Convergence}
\end{figure}

We then consider the equivalent of a $5M_{\earth}$ planet in our simulations to consider smaller planetary masses, with $5M_{\earth}$ being $0.25 M_\mathrm{iso}$ for the disc aspect ratio $h = 0.05$. In this case there are differences in the returned accretion rate values for dust with smaller Stokes numbers across the range of $f/\tau$. Therefore we instead took an average of these accretion rate values for the last 300 planetary orbits to reduce the effects of small fluctuations where the accretion rate as a function of $f/\tau$ is needed to make comparisons. We then plot the accretion rate as a function of $f/\tau$ for different Stokes numbers, $\mathrm{St} = 0.22, 0.017, 10^{-3}$, and $10^{-4}$, in Fig.~\ref{fig: 5ME Dust Factor Convergence} to investigate this variability. As a result it is clear that the choice of $f/\tau$ has less effect on the accretion rate for the larger dust grains, however, we can see the difference for the two smaller species. In this case we ran an additional simulation with a larger $f/\tau = \frac{100}{\pi}\Omega_p$, to check for convergence, and we find that the accretion rates do converge towards these larger values.

From this investigation we conclude that in this lower mass regime we require factors $f/\tau > \frac{10}{\pi}\Omega_p$ to get convergence in the dust accretion rate for the smallest grains. Since we also confirmed above that the dust accretion rate in the large planet mass regime is largely unaffected by the differing accretion factors, we opt for an accretion rate of $f/\tau = \frac{20}{\pi}\Omega_p$ for the dust that can be applied to all dust grains.   

\section{Gas Accretion Resolution Check}\label{Appendix: Gas Accretion Resolution Check}

Testing of the accretion routine across different resolutions is conducted in 3D to confirm it is robust independent of grid resolution. This is carried out by tracking the gas accretion rate across three different resolutions with grid sizes, using the grid resolution used throughout this work, $(N_r, N_\theta, N_\phi) = (512,60,1024)$, $1.5$ times this resolution with $(768,90,1536)$, and $2.0$ times this resolution with $(1024,120,2048)$. The accretion rate is tracked for $680$ orbits for the two coarser grids, while due to the prohibitively long run times, the highest resolution grid is only run for $\sim 400$ orbits. 

Despite the shorter run time for the highest resolution, this test confirms there is little variation across the computed accretion rates using our algorithm, with mean ratios between the high and standard resolution of $1.04$, between the high and mid resolution of $1.06$ and between the mid and standard resolution of $0.99$. With differences of only several percent between resolutions the standard resolution is therefore suffice for the purposes of this work. 

\begin{figure}
    \centering
    \includegraphics[width=\columnwidth]{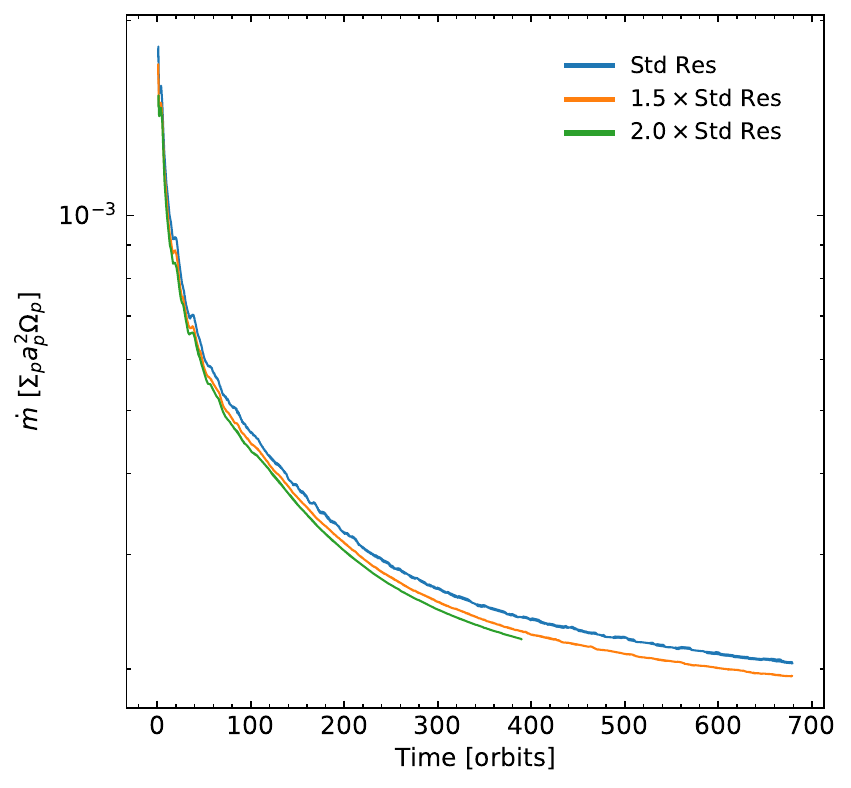}
    \caption{Gas accretion rate as a function of time in planetary orbits for simulations with three different resolution grids, the standard resolution (Std Res, blue) with $(N_r, N_\theta, N_\phi) = (512,60,1024)$, $1.5$ times this resolution with $(768,90,1536)$ (orange), and  $2.0$ times this resolution with $(1024,120,2048)$ (green). Due to the long timescales required, the highest resolution is only run for $\sim 400$ planetary orbits.}
    \label{fig: Gas Accretion Resolution Check}
\end{figure}


\bsp	
\label{lastpage}
\end{document}